\documentclass{aa}  

\usepackage{linenoaa}

\usepackage{graphicx}
\usepackage{txfonts}
\usepackage{booktabs}
\usepackage{titlesec}
\usepackage[dvipsnames]{xcolor}
\usepackage{subfigure}
\usepackage{pdflscape}
\usepackage{caption}
\usepackage{sidecap}   

\usepackage{color}
\usepackage{natbib,twoopt}
\usepackage[hyphenbreaks]{breakurl}
\usepackage[breaklinks]{hyperref}      
\bibpunct{(}{)}{;}{a}{}{,}             
\definecolor{cobalt}{rgb}{0.06, 0.2, 0.65}
\hypersetup{
  colorlinks,
  citecolor=blue,
  linkcolor=blue,
  urlcolor=blue,
}
\makeatletter
  \newcommandtwoopt{\citeads}[3][][]{\href{http://adsabs.harvard.edu/abs/#3}%
    {\def\hyper@linkstart##1##2{}%
     \let\hyper@linkend\@empty\citealp[#1][#2]{#3}}}
  \newcommandtwoopt{\citepads}[3][][]{\href{http://adsabs.harvard.edu/abs/#3}%
    {\def\hyper@linkstart##1##2{}%
     \let\hyper@linkend\@empty\citep[#1][#2]{#3}}}
  \newcommandtwoopt{\citetads}[3][][]{\href{http://adsabs.harvard.edu/abs/#3}%
    {\def\hyper@linkstart##1##2{}%
     \let\hyper@linkend\@empty\citet[#1][#2]{#3}}}
  \newcommandtwoopt{\citeyearads}[3][][]%
    {\href{http://adsabs.harvard.edu/abs/#3}
    {\def\hyper@linkstart##1##2{}%
     \let\hyper@linkend\@empty\citeyear[#1][#2]{#3}}}
\makeatother

\titleformat{\paragraph}
{\normalfont\normalsize}{\theparagraph}{1em}{}
\titlespacing*{\paragraph}
{0pt}{3.25ex plus 1ex minus .2ex}{1.5ex plus .2ex}

\newcommand{\bse}{BSE\xspace}
\newcommand{\mocca}{MOCCA\xspace}
\newcommand{\cmc}{CMC\xspace}

\newcommand{\nbodySeven}{nbody7\xspace}

\newcommand{\nbodySixppGPU}{nbody6++GPU\xspace}
\newcommand{\petar}{PeTar\xspace}
\newcommand{\msun}{$\rm M_{\odot}$\xspace}

\titlerunning{Double white dwarf binary population in MOCCA star clusters}
\authorrunning{L. Hellström et al.}

\begin{document}

   \title{Double white dwarf binary population in MOCCA star clusters}

   \subtitle{Comparisons with observations of close and wide binaries}

   \author{
    L. Hellström$^{1}$,
    M. Giersz$^{1}$, A Hypki$^{1,3}$, D. Belloni$^{2}$, A. Askar$^{1}$, G. Wiktorowicz$^{1}$
    }

   \institute{Nicolaus Copernicus Astronomical Center, Polish Academy of Sciences, ul. Bartycka 18, PL-00-716 Warsaw, Poland
   \\
              \email{hellstrom@camk.edu.pl}
         \and
             Departamento de Física, Universidad Técnica Federico Santa María, Av. España 1680, Valparaíso, Chile
             \and
   Faculty of Mathematics and Computer Science, A. Mickiewicz University, Uniwersytetu Pozna\'nskiego 4, 61-614 Pozna\'n, Poland
             }

   \date{Received 8 May 2024 / Accepted 18 July 2024}

  \abstract{There could be a significant population of double white dwarf binaries (DWDs)  inside globular clusters (GCs); however, these binaries are often too faint to be individually observed. We have utilized a large number GC models evolved with the Monte Carlo Cluster Simulator (MOCCA) code to create a large statistical dataset of DWDs. 
  These models include multiple-stellar populations, resulting in two distinct initial populations: one dense and the other less dense.
  Due to the lower density of one population, a large number of objects escape during the early GC evolution, leading to a high mass-loss rate. In this dataset we have analyzed three main groups of DWDs, namely in-cluster binaries, escaped binaries, and binaries formed from the isolated evolution of primordial binaries. We compared the properties of these groups to observations of close and wide binaries. We find that the number of escaping DWDs is significantly larger than the number of in-cluster binaries and those that form via the isolated evolution of all primordial binaries in our GC models. This suggests that dynamics play an important role in the formation of DWDs. For close binaries, we found a good agreement in the separations of escaped binaries and isolated binaries, but in-cluster binaries showed slight differences.  We could not reproduce the observed extremely low mass WDs  due to the limitations of our stellar and binary evolution prescriptions. For wide binaries, we also found a good agreement in the separations and masses, after accounting for observational selection effects. Even though the current observational samples of DWDs are extremely biased and incomplete, we conclude that our results compare reasonably well with observations.}


   \maketitle

%

\section{Introduction}

Globular clusters (GCs) are self-gravitating spherical collections of tens of thousands of stars  up to several million  stars. The cores of these clusters are very dense (up to ${\sim10^6}$ stars per pc$^3$), making stellar collisions and dynamical interactions relatively frequent. These interactions can affect not only the stars, but also the cluster as a whole. For a binary, an interaction with another star can lead to a hardening or softening, depending on the interaction. It can also lead to the formation of a whole new binary through an exchange or a merger in a three-body or four-body interaction. In some cases these interactions can even disrupt the binary so that it dissolves. During a three-body interaction, binaries can increase the kinetic energy of a third object or of the binary's center of mass.

Globular clusters are generally very old and without any ongoing star formation. This causes them to almost exclusively contain old stars and, combined with the fact that they are found in almost all galaxies, makes them very interesting targets for observations and theoretical studies. Because of this, several research groups around the world are working on codes to simulate the evolution of these clusters in order to better understand their evolution. Some examples of these codes include, but are not limited to, the \mocca code \citep{MOCCA2}, \cmc \citep{kremer2020,Rodriguez2021}, \nbodySixppGPU \citep{Wang2015,wang2016,kamlah2022,Arcasedda2024a}, \nbodySeven \citep{banerjee2022}, and \petar \citep{Wang2020}.  

White dwarfs (WDs) are relatively abundant in GCs \citep{torres2015}, and often make up a large portion of the stellar population. White dwarfs are important objects inside clusters since they can, among other things, be used to determine the age of their host cluster  \citep[e.g.,][]{whiteDwarfCosmochronology}. The cooling process of WDs is believed to be quite well understood, so observations of their temperature and luminosity can be used to determine the age of the WD, and thus the cluster. However, recent developments in understanding the physical processes occurring during their cooling have been challenging this picture \citep[e.g.,][]{Bedard2024} 

There are several different populations of WDs. The first and most basic one is that composed of single WDs. Another sizable population is made up of detached binaries hosting main-sequence (MS) stars. In this case, the WD and the MS star have a wide enough orbit so that each star does not fill its Roche lobe, and thus there is no ongoing mass transfer \cite{campos2018}. The semi-detached counterpart of these systems is called a cataclysmic variable. A cataclysmic variable is a binary with a WD and a Roche-lobe filling secondary star, typically a MS star. These binaries are very tight with a short period (${\sim1-6}$\,h) such that the secondary fills its Roche lobe and transfers mass onto the WD. This mass transfer typically results in an accretion disk around the WD, which can often   be seen in ultraviolet and X-ray observations. The last population of WDs we consider are the focus of this paper, and correspond to double WD binaries (DWDs), which consist of two WDs.

Double WD binaries can, in general, form in two ways: the progenitors to the WDs were formed together in a binary system \citep{Belloni2023} or the binary was formed during a dynamical interaction \citep{kratter2011}. The resulting DWD can have different properties depending on the formation mechanism. The properties of primordial DWDs, that is DWDs forming from the evolution of primordial binaries, depend on the initial binary parameter distribution of primordial binaries.
The eccentricity distribution of a population of primordial binaries, by definition and according to  our setup, has a thermal distribution \citep{jeans1919}, while a population of dynamically formed binaries has higher eccentricities than a thermal distribution since they experience only a few dynamical interactions that will, on average, increase their eccentricities. Meanwhile, the semi-major axis distribution of primordial binaries  depends on the initial distributions, while for dynamically formed binaries it depends on the properties of the dense environment and gravitational encounters in which the binary is formed.  In addition, due to the dense environments where DWDs are formed, it is likely that the binary can eventually experience more dynamical interactions that may affect the semimajor axis further.

Since DWDs can be observed across the electromagnetic spectrum and with planned gravitational wave detectors, they are known as multi-messenger sources. Both ground- and space-based telescopes and observatories can detect and measure WDs by electromagnetic radiation. Double WD binaries can emit radiation in a wide range of wavelengths, from optical to X-rays. In addition, with more precise and sensitive gravitational wave (GW) detectors, it will be possible to detect DWDs through the GW radiation they emit \citep{Liu2099, Maselli2020, Carvalho_2022, LISApaper}. Finally, there are speculations that neutrinos produced in the core of WDs can be detected and used as probes to study the interiors of the WDs \citep{Drewes_2022}.

Recent studies have indicated that massive WDs most likely dominate the innermost regions of core-collapsed GCs such as NGC 6397 \citep{kremer2021}. These populations, through dynamical interactions, stop the cluster from completely collapsing. This produces very tight, inspiraling WD binaries that may be observed with future telescopes.

This paper  focuses on DWDs from simulations of GCs with multiple-stellar populations \citep{Lee1999, gratton2012, bastian2018}. We   focus on in-cluster, escaped, and isolated binaries (i.e., binaries with the same properties as those inside the clusters, but evolved in isolation without dynamical interactions or relaxation affecting them). These binaries are then compared to observations of DWDs in two regions; close binaries where the data is retrieved from \cite{brown2020}, \cite{kosakowski2023}, and \cite{schreiber2022} and wide binaries where the data is retrieved from \cite{elbadry2021} and \cite{heintz2022}. This was done to get a better understanding of how realistic the DWD populations are in clusters from our simulations. We also wanted to find out if there are differences between the DWD populations inside and outside of clusters (escapers and binaries evolved in isolation). Finding any potential differences would be an important discovery since it could aid in determining the origin of observed WD binaries and help our understanding of whether the field population of DWDs can be attributed to escapers from clusters. This also relates to an upcoming study where we will investigate the GW signals from WD binaries: if we   find differences in these populations, then we can also expect to see differences in their respective GW signals. This comparison, however, proved to be more difficult than we expected; due to severe limitations in the parameter space of observations, we had to limit our data greatly in order to make a proper comparison. Nevertheless, when restricting our data in ways that agree with observational techniques and limitations we find good agreement everywhere, except for the masses of close DWDs. 

\section{Observational sample}
\label{sec:obsSample}
\subsection{Close binaries}
The data for close DWDs were obtained from \cite{brown2020}, \cite{kosakowski2023}, and \cite{schreiber2022}. \cite{brown2020} uses the Extremely Low Mass  (ELM) Survey  to find a sample of 98 detached, close double WD binaries. This survey targeted $< 0.3$ M$_{\odot}$ helium-core WDs binaries with periods between $0.0089$ and 1.5 days. The absolute magnitude in the G band of the faintest object in this sample is $11.24$ mag. These WDs are so close to each other that they cannot be resolved separately. Therefore, this magnitude limit is for the two WDs combined. \cite{kosakowski2023} is a continuation of the ELM Survey where they found 28 additional DWDs. One important point to mention is that ELM WDs are rare; \cite{obrien2024} found 54 candidate DWDs, but none of them can be considered ELMs. Unfortunately, \cite{obrien2024} does not include periods for any of these close binaries. Instead, we add another sample of higher mass, close DWDs from \cite{schreiber2022} where 57 close DWDs were collected from different surveys and there was no limitation to only ELMs.

\subsection{Wide binaries}
For wide binaries we used the published catalog from \cite{heintz2022}, which contains data of 1590 wide DWDs from \textit{Gaia} EDR3 \citep{gaiaEDR3} and \cite{elbadry2021}. The goal of \cite{heintz2022} was to check existing methods for determining WD ages. They used wide DWDs, separated by more than 100 au, and made the assumption that at these distances, the stars would evolve independently without any significant interactions between the two binary components. In addition, they assumed that the stars in the binary were born from the same molecular cloud at approximately the same time.

This dataset includes a chance alignment factor ($R\_chance\_align$), which determines the chance that the detected binary is in fact not a binary but two single stars that, due to chance, appear to be bound in a binary. This produces an upturn in the tail at larger separations. Section 3 of \cite{elbadry2021} explains this in detail. We use this chance alignment factor and remove all binaries with $R\_chance\_align > 0.1$, which leaves us with 1389 binaries. 

The apparent magnitudes of the WDs in the \textit{Gaia} G band, $G$, are limited to 13 mag < G < 21 mag; however, this is not a cut done by the authors, but a limitation on observational instruments. Using the parallax, $p$, we can estimate the absolute magnitude with 

\begin{equation}
    G_{abs} = G + 5 - 5\log\left(\frac{1}{p}\right)     
.\end{equation}
Using the parallaxes and apparent G-band magnitudes from \cite{heintz2022} we estimate that the faintest WD in this dataset will have an absolute magnitude in the G band of $\sim$16 mag. In addition to this, due to the finite age of the Galactic disk, there is a peak in the white dwarf luminosity function at $\sim$16 mag \citep{garciaBerro2016}. Thus, we remove from our numerical data all binaries where either member has a magnitude above this limit.

Figure 7 in \cite{heintz2022} shows that in their dataset, for a large number of binaries, the cooling age of the primary, the more massive star, is shorter than the secondary. This is not expected in isolated binaries, and might indicate that these systems used to be triple systems where the inner binary interacted through mass transfer or a merger, and that this inner binary was later split or merged, and thus formed a binary with the third object. It is also possible that a  nonmonotonic initial-to-final mass relation can cause this \citep{marigo2020}. These objects were removed from our comparisons.

\section{Numerical simulations}
\label{sec:numericalSimulations}
\subsection{The \mocca code}\label{sec:method}
   We use the MOnte Carlo Cluster simulAtor, \mocca\footnote{\url{http://moccacode.net}} \citep{MOCCA1,MOCCA2}, and more specifically the simulations from MOCCA-SURVEY Database III \citep{hypki2024}, which is an upgrade over the MOCCA-SURVEY Database II \citep{hypki2022} and MOCCA-SURVEY Database I \citep{moccaSurvey1}. \mocca is a GC simulator built on a Monte Carlo approach. It allows the usage of multiple stellar populations with different properties, provides greater control over initial cluster conditions, and allows the quick evolution of large clusters with more than one million initial members.
   
   An important aspect of the \mocca~code for our purposes is how stellar and binary evolution is treated, which is done with the modified \bse~code \citep{Hurley2000,Hurley2002}.
The \bse~code consists of a set of algorithms describing single star evolution, from zero-age main-sequence stars to later stages of stellar evolution, and binary evolution, taking into account angular momentum loss mechanisms, different modes of mass transfer, and tidal interaction.
\bse~has been widely used to investigate different astrophysical objects and is characterized by its generally high level of accuracy in the analytic fitting formulae on which it is based.

Since the publication of \bse, there have been several upgrades to the code \citep[see, e.g.,][for more details]{kamlah2022}. The main upgrades that are relevant for this study are the inclusion of a proper prescription for cataclysmic variable evolution \citep{belloni2018b,belloni2019} and the inclusion of improved wind prescriptions \citep{belczynski2016}.

There are, however,   two main problems in \bse~that play an important role in the types of binaries we are interested in.
The first  is related to WD evolution, which is not properly handled by \bse.
The luminosity evolution of WDs is modeled in \bse~using the \citet{Mestel_1952} standard cooling theory.
This theory though does not take into account important properties such as core composition and surface abundances as well as crucial features such as crystallization, which can significantly delay the WD cooling for gigayears and has been shown to be very important for explaining several features in WD populations \citep[e.g.,][]{Tremblay_2019,schreiber_2021,Bagnulo_2022,Blatman_2024}.

We therefore improved the WD evolution in our analysis by computing more accurate WD properties via interpolation and extrapolation using the cooling sequences calculated by \citet{Bedard2020}, assuming a thin hydrogen atmosphere. We assumed all WDs have thin layers, but checked the impact of the assumed thickness of the hydrogen layer by assuming all have thick layers and obtained very similar magnitudes. The impact was not large enough to affect the results.
Provided the WD age and mass, we linearly interpolate--extrapolate through the sequences to obtain the WD properties, such as effective temperature, radius, and luminosity.
These sequences were computed assuming solar metallicity, which is higher than assumed in our GC simulations.
However, the impact of metallicity on WD evolution regarding the properties in which we are interested is virtually negligible \citep{Renedo2010}.
That said, for the purposes of our analysis it is reasonable to use the solar metallicity sequences calculated by \citet{Bedard2020}.
The main motivation to use such sequences among the few that are available comes from the fact that they were used by \citet{heintz2022}.

\subsection{Initial conditions}
\label{sec:mocca_data}
We used 197 cluster simulations with a wide range of initial cluster parameters. A full description of the different cluster initial conditions can be found in \citet{hypki2024}. The most important parameters are the following:
\begin{itemize}
    \item many different combinations of sizes of clusters, from 550 thousand to 2.6 million initial objects (two multiple stellar populations);
    \item both tidally filling and underfilling clusters;
    \item both a 10\% and 95\% initial binary fraction;
    \item a Kroupa mass function between 0.08 and 150 M$_{\odot}$;
    \item different Galactocentric distances;
    \item different properties of multiple populations.
\end{itemize}

A big difference with these simulations compared to earlier simulations is that we are using multiple-stellar populations in the framework of the  asymptotic giant branch scenario \citep{calura2019} in these new simulations, and thus a new path to GC evolution. Instead of having one tidally underfilling dense cluster, we have two populations: a tidally filling (or slightly tidally underfilling) less dense population with a higher maximum initial stellar mass, and a tidally underfilling denser population. This leads to a large number of early escapers from the first population where approximately 30-40\% of mass is removed in the first few megayears. These binaries are mostly primordial and undisturbed from dynamical interactions since they escape at an early stage before they have had time to interact with other objects. The second population is  denser, but the maximum stellar mass is lower; there is no formation of massive stars in this population. This limit is set to 20 M$_{\odot}$, which means that there are no supernovae forming BHs from the second population. The previously used initial cluster conditions are quite similar to the second population, while the first population is not as tightly bound and is able to lose a large amount of mass early in the evolution.

The choice of a 10\% or 95\% binary fraction affects the initial binary properties. For example, in a 10\% binary fraction cluster, the largest semi-major axis value is 100 au and the distribution is uniform in log(a). With a 95\% binary fraction, we allow binaries to form with larger separations and the initial distribution of semi-major axis and eccentricity is based on \cite{kroupa1995} and \cite{belloni2017}. Since these binaries are often disrupted by dynamical interactions, a 95\% binary fraction has been found to agree well with observations \citep{belloni2017,belloni2018a}. For both binary fractions, the initial eccentricity distribution is thermal.

\subsection{Star cluster and isolated binary evolution}
\label{sec:clusterAndBinaryEvolution}
For each initial setup, we ran two simulations, one with dynamical interactions between objects and one run with no dynamical interactions (i.e., pure binary or stellar evolution where all objects in our initial star cluster model are evolved in isolation). This allows us to understand the impact of dynamics on the primordial binary population and DWD formation. In addition, whenever an object is ejected from a cluster, we used \bse to continue the evolution of this object. Furthermore, we split up primordial and dynamically formed binaries in the cluster simulation with dynamics. This gives the following four groups:
\begin{itemize}
    \item cluster-primordial: DWDs with progenitors born together found inside an evolving cluster;
    \item cluster-dynamical: DWDs that were formed in a dynamical interaction found inside an evolving cluster;
    \item standalone evolution: DWDs evolved in isolation;
    \item escapers:  DWDs that escaped from the cluster simulations, and that evolved as isolated binaries after escape.
\end{itemize}

These simulations are run up to 15 Gyr in order for us to obtain data from a wide range of evolutionary stages at different times.

\subsection{Further calculations}
\label{sec:furtherCalc}
There is evidence for natal kicks on WDs \citep{davis2008, hamers2019} that could lead to the destruction of fragile wide binaries. The magnitude of these kicks is believed to be on the order of $\sim 0.75$ km/s \citep{elbadry2018}. This is not included in our simulations, and thus we add the energy related to a kick of $0.75$ km/s for each star to the binding energy of the binary. This can be seen as an upper limit scenario. We check whether this causes the binary to dissolve. If it does, we remove this binary from the sample. If not, we calculate the new semi-major axis related to the new binding energy. 

To calculate the absolute magnitude of the WDs from \mocca we use the \textsc{FSPS}\footnote{\url{https://github.com/cconroy20/fsps}} code \citep{conroy2009, conroy2010}. For one binary, this code can calculate the absolute magnitudes of both components in a binary using the mass, radius, luminosity, and effective temperature. This provides us with absolute magnitudes for different instruments and filters, such as the \textit{Gaia} G band.

\subsection{Observational selection effects}
\label{sec:observationalSelection}
Binary stars can be detected through different methods depending on the distance between the two components. Close binaries are detected, for example, through  photometric surveys, spectroscopic observations, or X-ray observations, while wide binaries
are too far apart for us to reliably use photometric data. Instead, if two
wide separated stars have a common proper motion across the sky or similar radial velocities, they can be classified as a binary candidate, and in more detailed follow-up observations this can be confirmed. Due to the different methods of obtaining observational data, there are also different limitations. We need to consider these observational selection effects when comparing our data. Thus, we apply different filtering criteria on the close and wide binaries.
These criteria are explained in the following sections.

\subsubsection{Close binaries}
\label{sec:closeFiltering}
For close binaries, we use the following filtering criteria on our \mocca data:
\begin{itemize}        
    \item P < 1.5 day;
    \item $G_{\mathbf{abs}}$ < 11.24 mag, where $G_{\mathbf{abs}}$ is the combined absolute magnitude of the two WDs:
    \begin{equation}
        G_{\mathbf{abs}} = -2.5 \log(10^{-0.4G_{\mathbf{abs},1}} + 10^{-0.4G_{\mathbf{abs},2}}),
    \end{equation}
    where $G_{\mathbf{abs},1}$ and $G_{\mathbf{abs,}2}$ are the absolute magnitudes in the G band of the primary and secondary WD.
\end{itemize}
We did not use any filtering based on WD mass.

\subsubsection{Wide binaries}
\label{sec:wideFiltering}
For wide binaries, we use the following filtering criteria on our \mocca data:
\begin{itemize}
    \item semi-major axis: $a > 100$ AU;
    \item absolute magnitude: $G_{\mathbf{abs}} < 16.1$ mag;
    \item addition of kinetic energy corresponding to kicks of WDs to the binding energy of the binary (see Sect.~\ref{sec:furtherCalc}).
\end{itemize}
As an additional step we use the total ages reported in \cite{heintz2022}, limited to $T_{age} < 10$ Gyr, and extract a subset of our data with the same age distribution. This gives us a dataset with WDs with a wide range of ages. This filtering is only used for the results presented in Sect.~\ref{sec:results_WideHeintzAges}.

\subsection{Projection}
\label{sec:projMethod}
It is often impossible to determine the semi-major axis of wide binaries from observational data. Instead, the projected separation between the two binary components is the parameter that can be determined. This means that to compare the binaries from our simulations to the observed binaries, we need to project our semi-major axis. The projection method we use is derived from chapter 2 of \cite{hilditch_2001}. The radius vector (r) joining the two binary components at a given time is at an angular position $\theta$, the true anomaly relative to the position of periastron of the orbit. This radius vector can be projected onto the x and y coordinate axes as
\begin{equation}
    \begin{aligned}
        x = r \cos \theta [\cos \Omega \cos \omega - \sin \Omega \sin \omega \cos i] + \\
        r \sin \theta [-\cos \Omega \sin \omega - \sin \Omega \cos \omega \cos i]\\
        y = r \cos \theta [\sin \Omega \cos \omega + \cos \Omega \sin \omega \cos i] + \\
        r \sin \theta [-\sin \Omega \sin \omega + \cos \Omega \cos \omega \cos i],
    \end{aligned}
\end{equation}where $\Omega$ is the longitude of the ascending node, $\omega$ is the longitude of periastron, and $i$ is the inclination of the orbit (see Fig.~2.5 in \citealt{hilditch_2001}). Knowing the masses, semi-major axis, and eccentricity of the binary system, we can randomly sample the Keplerian orbital elements in Eq. 3 and the mean anomaly in order to use Kepler's equation to find the corresponding eccentric anomaly and then the true anomaly. This allows us to  obtain the Cartesian position coordinates of the two binary components. It is assumed that the observers line of sight is along the z direction, and thus with the position of the binary components we calculate the projected separation as 

\begin{equation}
    \mathbf{a_{proj} = \sqrt{(x_2 - x_1)^2 + (y_2 - y_1)^2}}
,\end{equation}where $x_n$ and $y_n$ are the x and y positions of the two respective binary components. To obtain an average projected separation, we do this 10000 times for each binary and use the average separation (for more details about this procedure, see \citealt{askar2018}).

\section{Results}
\label{sec:results}
We start the presentation of our results by comparing the projected separation (calculated according to Sect.~\ref{sec:projMethod}) with the mass of the WDs. For the observed close binaries we only consider the mass of the lower mass WD since the higher mass WD is typically not visible and the errors on the masses are usually very large. For all other datasets, both observations and simulations, we consider the mass of both WDs. Since observed close binaries cannot be resolved individually, there is no observational data for the separation of these binaries; however, we know the period. Using the periods, we assume that the eccentricity of these binaries is 0 due to circularization, and calculate the semi-major axis using Kepler's third law.

\begin{figure*}[ht]
    \centering
    \begin{minipage}[b]{12cm}
        \centering
        \includegraphics[width=\textwidth]{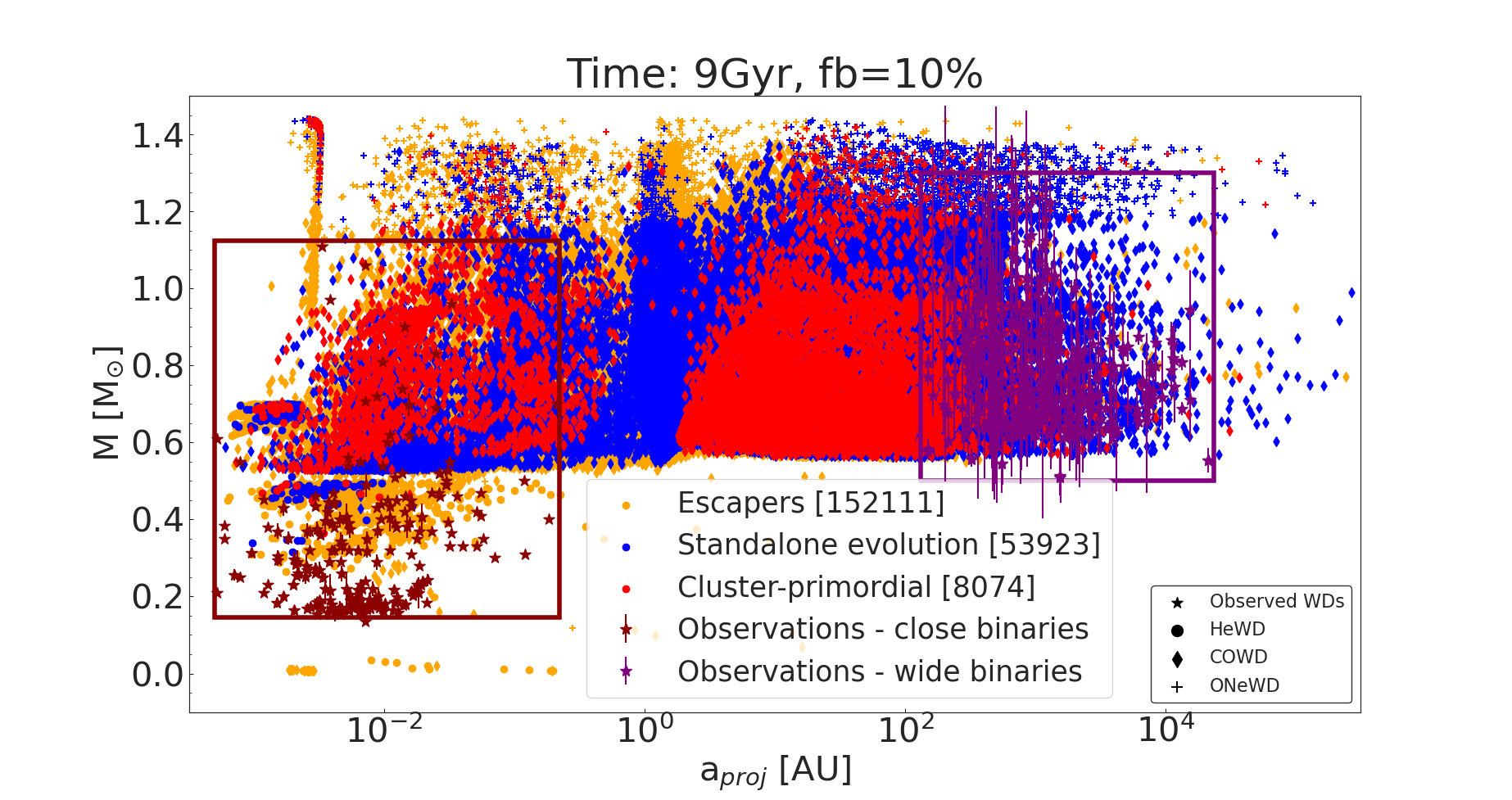}\\[2ex]
        \includegraphics[width=\textwidth]{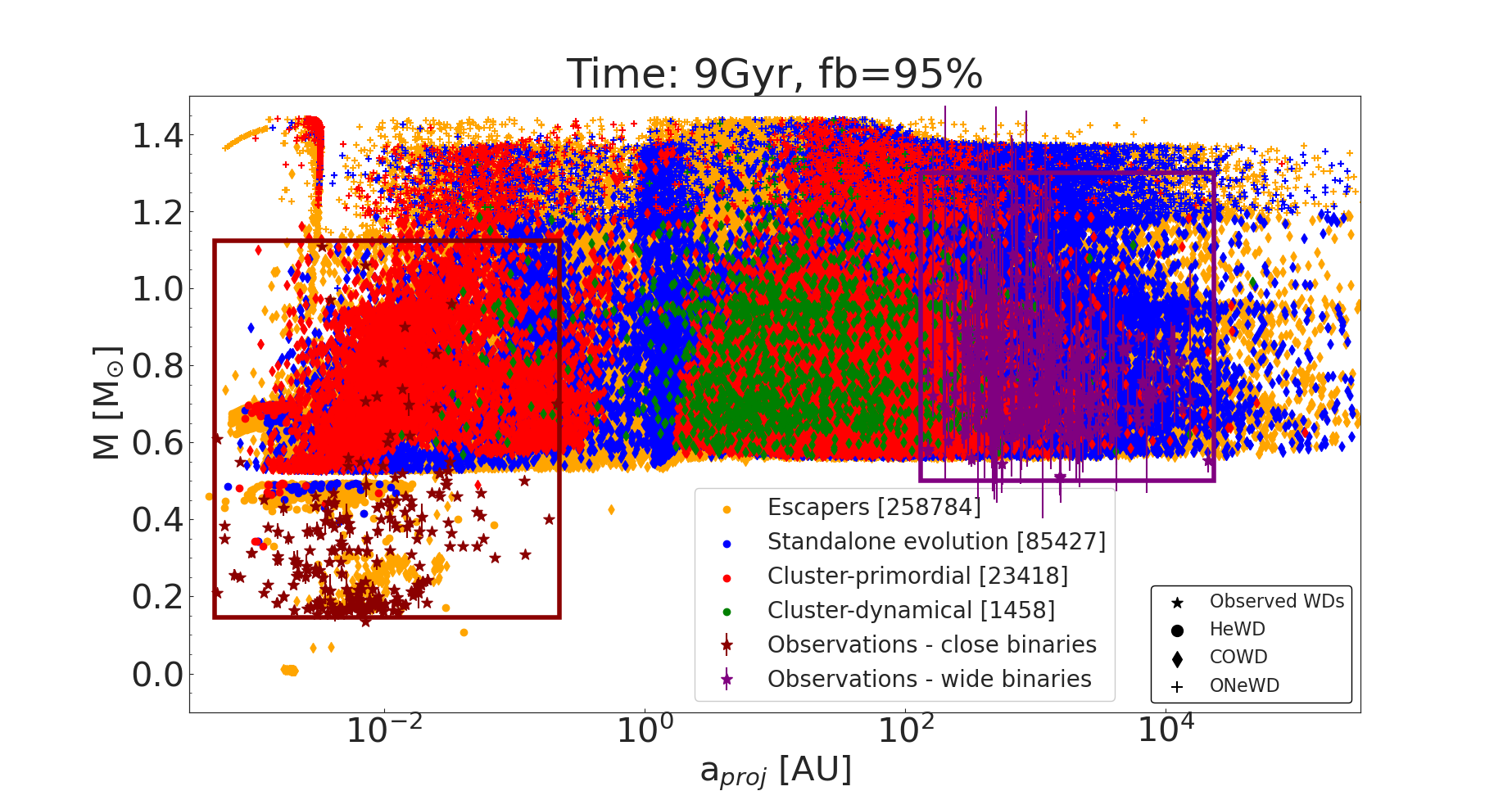}\\
    \end{minipage}
    \hspace{1em}  
    \begin{minipage}[b]{5cm}
        \caption{Projected separation plotted against mass of the lower mass WD in the binary for 10\% (panel a) and 95\% (panel b) initial binary fraction for the whole parameter space.  The numbers inside the square brackets in the legends show the total number of WD binaries in the runs. The different colors correspond to the different datasets: orange for escapers, blue for standalone binaries, red for in-cluster primordial binaries, green for in-cluster dynamically formed binaries, dark red for observations of close binaries, and purple for observations of wide binaries. In addition, we have four different types of markers: stars for observed WDs, circles for helium WDs, diamonds for carbon oxygen WDs, pluses for oxygen neon WDs.}
        \label{fig:aMass_010}
    \end{minipage}
\end{figure*}

In Fig.~\ref{fig:aMass_010} we show the projected separation plotted against the mass of the WDs for clusters with a 10\% and 95\%  initial binary fraction at 9 Gyrs snapshot. This plot shows the whole range of separations, from close to wide, for the four different groups defined in Sect.~\ref{sec:clusterAndBinaryEvolution}. 

The differences in the overall picture are very small between a 10\% and 95\% binary fraction, and thus we  discuss them together unless stated otherwise. Our magnitude limits are based on observational limitations, and thus we can only apply these limitations to separations where DWDs have been observed. In addition, the magnitude limits are different for close and wide binaries, which means that  we are unable to filter based on magnitude for these two plots. In later sections we   present the  10\% and 95\% binary fraction separately and we  include the magnitude limits discussed in Sect. \ref{sec:observationalSelection}.

A noticeable feature of the in-cluster binaries is a gap in separations between 0.6 AU and 25 AU. This gap only exists for in-cluster binaries, while for   escapers and isolated binaries many systems   occupy  this region in the parameter space. The reason for this gap is dynamics; more specifically, the threshold between soft and hard binaries lies in this gap. Binaries below this threshold will, on average, become harder due to dynamical interactions while binaries above will, on average, become softer as a result of dynamical interactions. This pushes binaries away from the threshold and creates this gap. This is one example of how important dynamical interactions can be for binaries in these dense environments. 

The distribution of escapers is very similar to that of  isolated binaries on a statistical scale. However, a very noticeable trend is that for all times and across all GC models we have far more escaped DWDs than isolated DWDs. The exact number depends on the initial conditions and the evolution time, but we found that usually the ratio is between 1 and 3. This seems to indicate that dynamics play an important role in the formation of these binaries, and without dynamics the formation of these binaries is not as effective as when dynamical interactions are included. 

There is also a big difference in the number of escapers and in-cluster binaries. The reason for this is most likely due to stellar evolution leading to rapid mass loss from the evolution of massive stars in the first 100 Myr of the cluster evolution. When the stars are initially evolving and losing mass the tidal radius   decreases strongly and rapidly. This means that many binaries are suddenly outside of the tidal radius, and are thus counted as escapers. The binaries can later evolve into DWDs. We discuss this in more detail in Sect.~\ref{sec:mocca_data}.

For very small separations, there is a hook-shaped formation, particularly for escapers and in-cluster binaries. This is related to cataclysmic variables that have later evolved into a DWD system. The fact that we have far fewer systems in this formation for isolated binaries once again shows the importance of dynamical interactions leading to their formation. For escapers in the 10\% initial binary fraction datasets, there are also a few lines going  toward smaller separations. These are also related to cataclysmic variables.

The observations of DWDs are currently limited to either spectroscopic binaries or wide binary DWDs. While the \mocca data spans a wide range of separations and masses, the  observations are either for very close  or for very wide binaries. We can clearly see that the observations are only including a very small part of the whole population. Because of this, we cannot make a direct comparison between the \mocca data and observations. We split our results into close and wide binaries and restrict our \mocca data, as explained in Sects. \ref{sec:closeFiltering} and \ref{sec:wideFiltering}. The results for close binaries are found in Sect.~\ref{sec:results-close} and for wide binaries in Sect.~\ref{sec:results-wide}.

\subsection{Close binaries}
\label{sec:results-close}
In order to go into more detail when looking at the close binaries, we isolate binaries according to the criteria described in \ref{sec:closeFiltering}, and   thus only focus on a small subset of the total population. We plot the projected separation against mass of the lower mass WD in Figs.~\ref{fig:perMass_fb0.1} and \ref{fig:perMass_fb0.95}. We only include the mass of the lower mass WD since, due to the observational technique, the errors on the mass of the higher mass WD are often very large. Our limit on magnitude means that over time we see that higher mass WDs are not visible, due to the cooling process of WDs. Already at 5 Gyr, all WDs with masses higher than $\sim0.55$M$_{\odot}$ are now undetectable. 

\subsubsection{Initial binary fraction of 10\%}
\begin{figure*}[ht]
    \centering
    \begin{minipage}[b]{12cm}
        \centering
        \includegraphics[width=\textwidth]{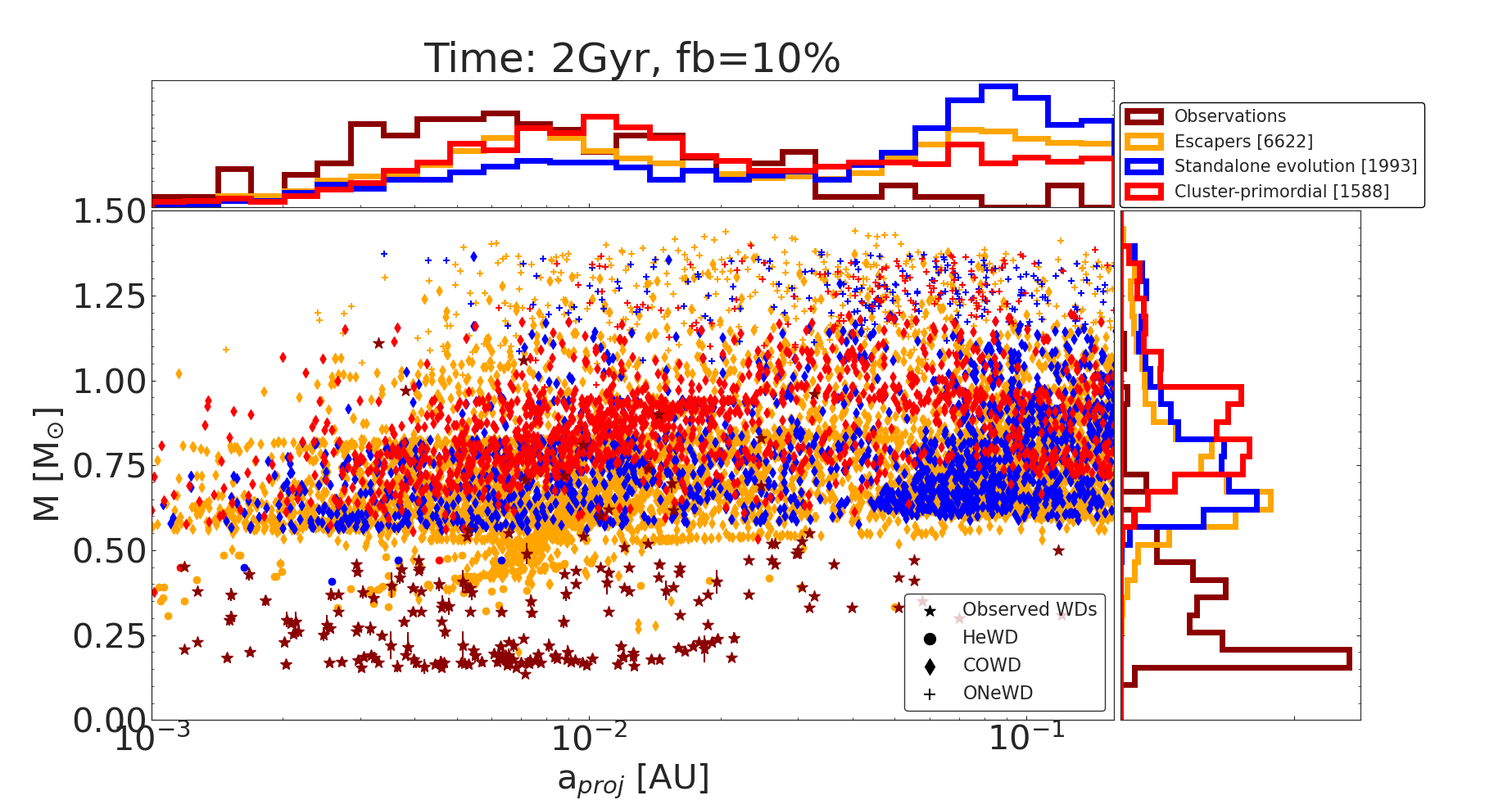}\\[2ex]
        \includegraphics[width=\textwidth]{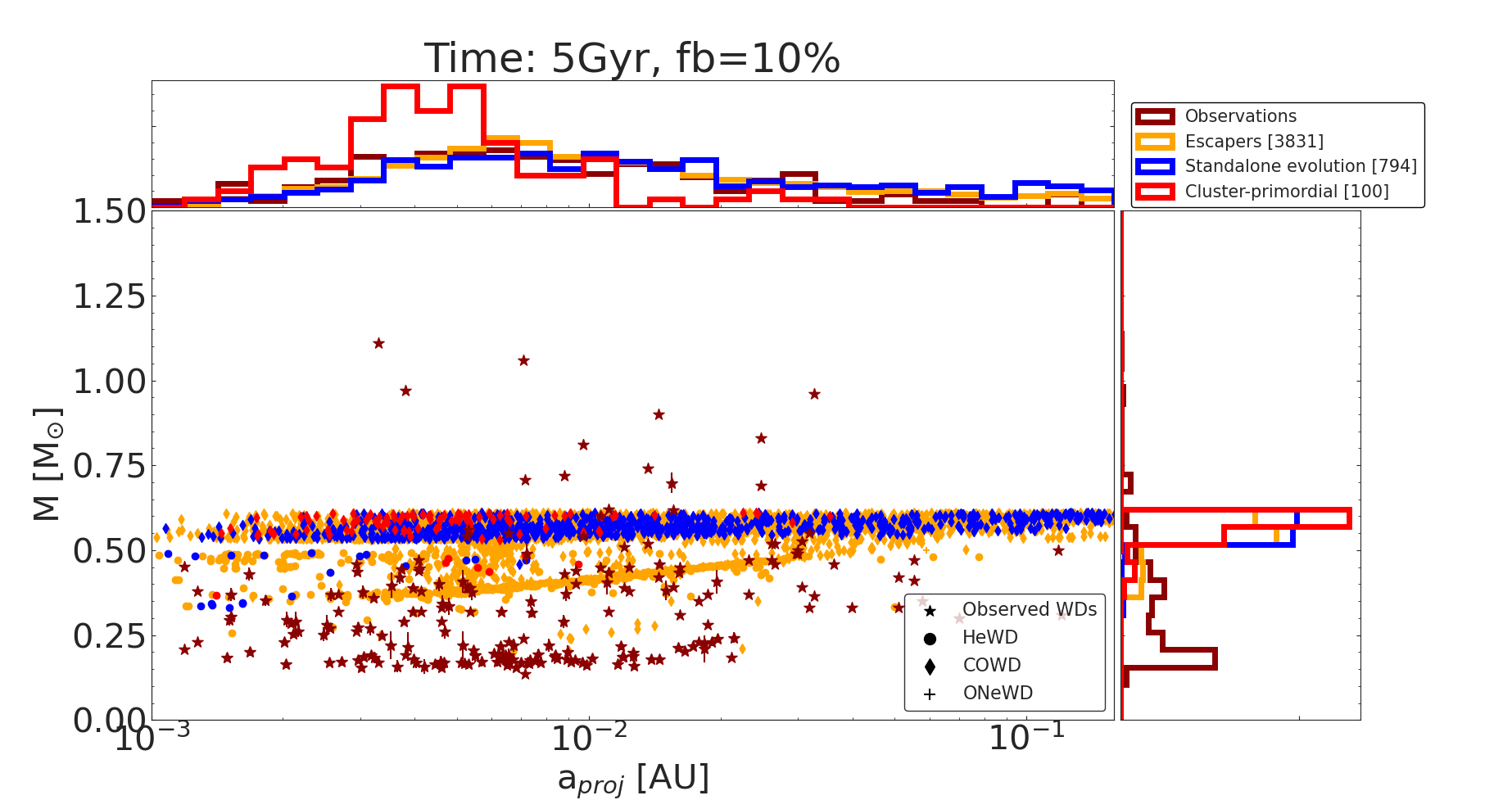}\\[2ex]
        \includegraphics[width=\textwidth]{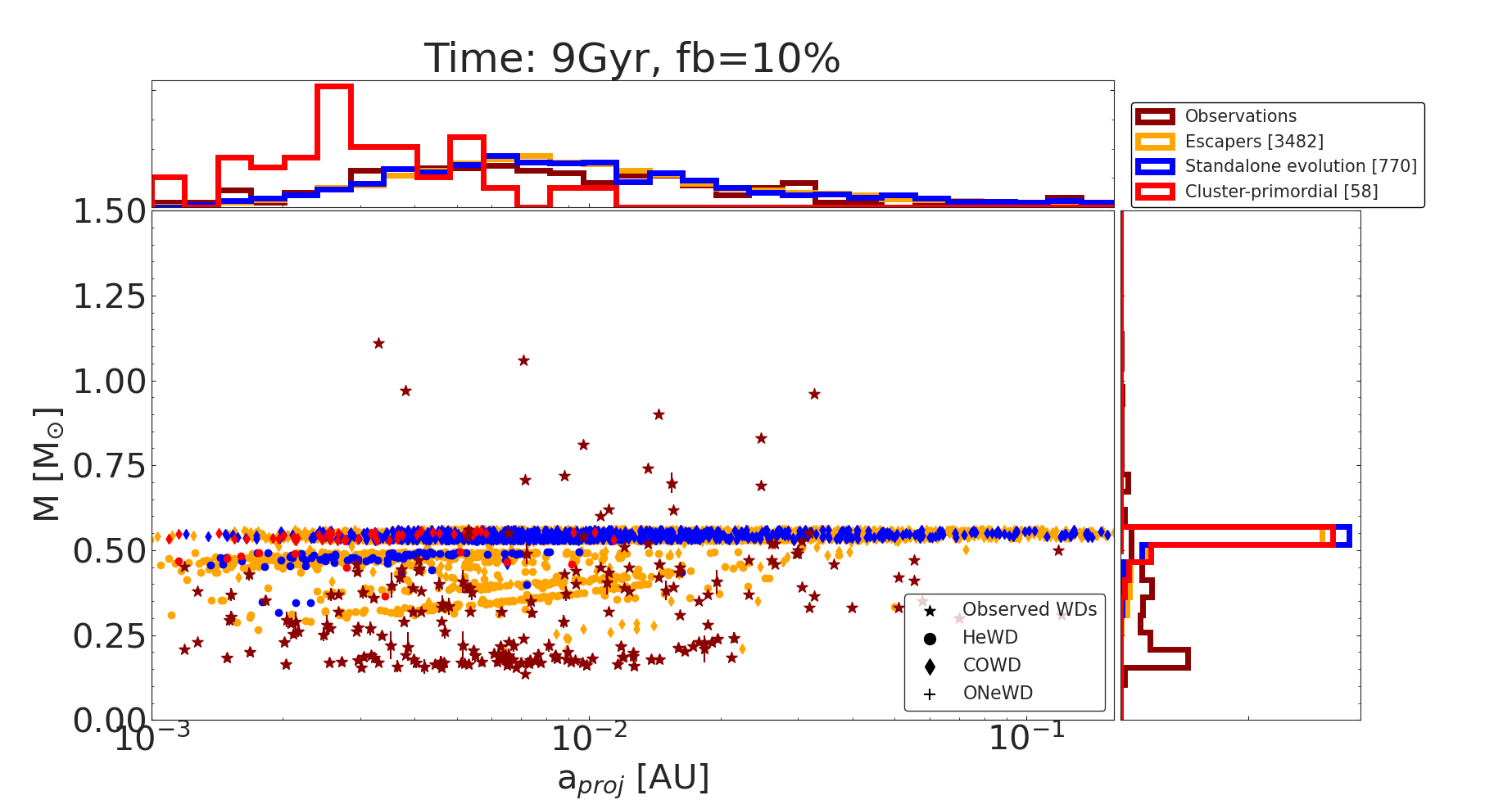}
    \end{minipage}
    \hspace{1em}  
    \begin{minipage}[b]{5cm}
        \caption{Projected separation plotted against mass of the lower mass WD in the binary for close binaries inside clusters with 10\% initial binary fraction. See Fig.~\ref{fig:aMass_010} for an explanation of panels, markers, and colors.}
        \label{fig:perMass_fb0.1}
    \end{minipage}
\end{figure*}

We  start by looking at the 10\% initial binary fraction plots in Fig.~\ref{fig:perMass_fb0.1}. At 2 Gyr the separations from all \mocca datasets are slightly too wide and the masses too high. At 5 Gyr, escapers and isolated binaries agree very well with the observations in separation. The in-cluster binaries at this time are biased toward smaller separations than the other datasets. At 9 Gyr, not much changes compared to 5 Gyr: the maximum mass is slightly lower because more binaries are nondetectable due to increasing magnitude as they cool down. However, for all snapshots, the masses are not well reproduced. We are not able to produce very low mass WDs with mass below 0.2 M$_{\odot}$ in our simulations. Instead, our close WDs usually have a mass between 0.25 and 0.4 M$_{\odot}$. There are a few observed WDs in this mass range as well, but the large majority are lower mass. 

These ELMs cannot form through common envelope evolution, which produces a problem for our comparisons. Currently, \bse does not handle this kind of evolution well. There is a large uncertainty in the magnetic braking, mass stability criteria, and mass transfer modeling. More precisely, to form these objects we need dynamically stable mass transfer \citep[e.g.,][]{belloni2023carb}. During this stable mass transfer involving a Roche-lobe filling subgiant or an unevolved red giant star, due to strong magnetic braking, the evolution is convergent and the binary evolves toward shorter orbital periods. This kind of mass transfer is not currently handled by \bse or any other population synthesis code. In order to be able to properly reproduce the distribution of mass for these kinds of ELMs, upgrades to our \bse are needed; however, this  is outside  the scope of this paper.

Escaping binaries are able to form lower mass WDs than in-cluster or isolated binaries, which can be seen as a subpopulation between 0.3 and 0.45 M$_{\odot}$. This is most likely due to dynamical interactions before they escaped their host cluster. Either the in-cluster binaries that would form these low mass WDs are  escaping or dynamical interactions are causing an exchange, a break-up, or a merger.

At $\sim0.51$ \msun we can see a soft limit for WD masses. There are WDs with masses lower  than this, but not many. This is a limit in \bse and a lower mass would require processes that are neither in \bse nor solved in theory. This includes, as mentioned above, uncertainty in magnetic braking, mass stability criteria, and mass transfer modeling.

\subsubsection{Initial binary fraction of 95\%}
\begin{figure*}[ht]
    \centering
    \begin{minipage}[b]{12cm}
        \centering
        \includegraphics[width=\textwidth]{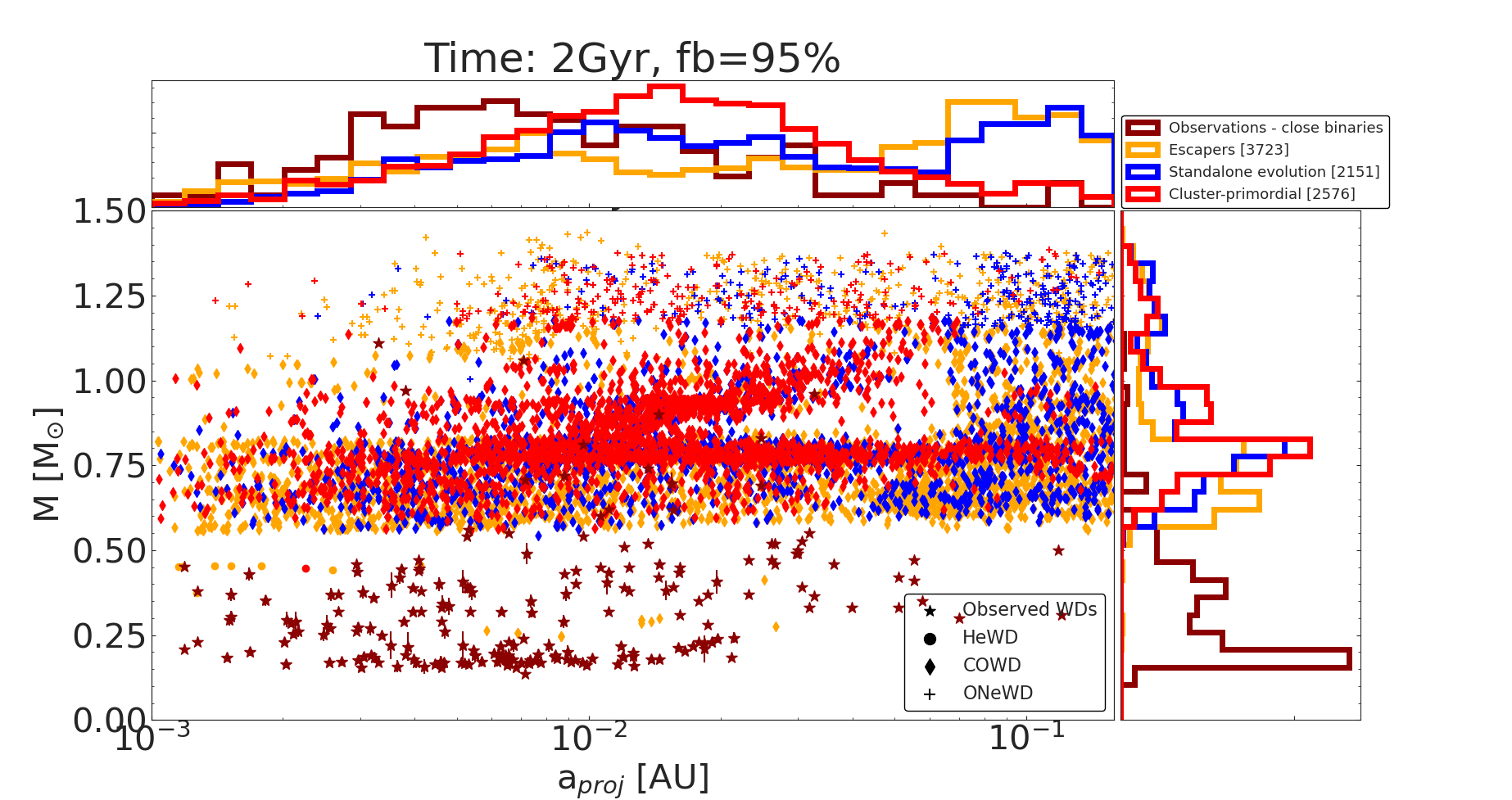}\\[2ex]
        \includegraphics[width=\textwidth]{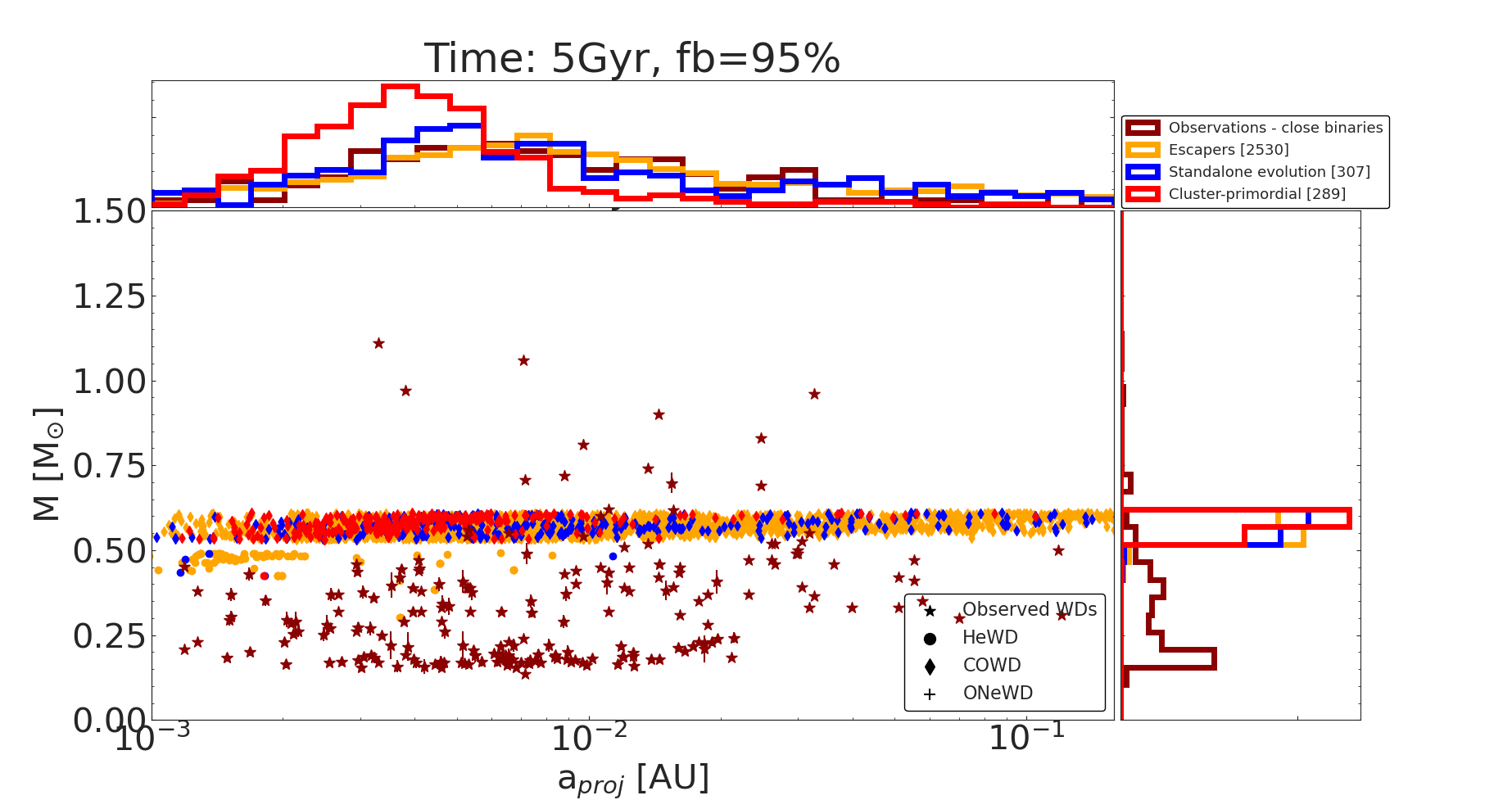}\\[2ex]
        \includegraphics[width=\textwidth]{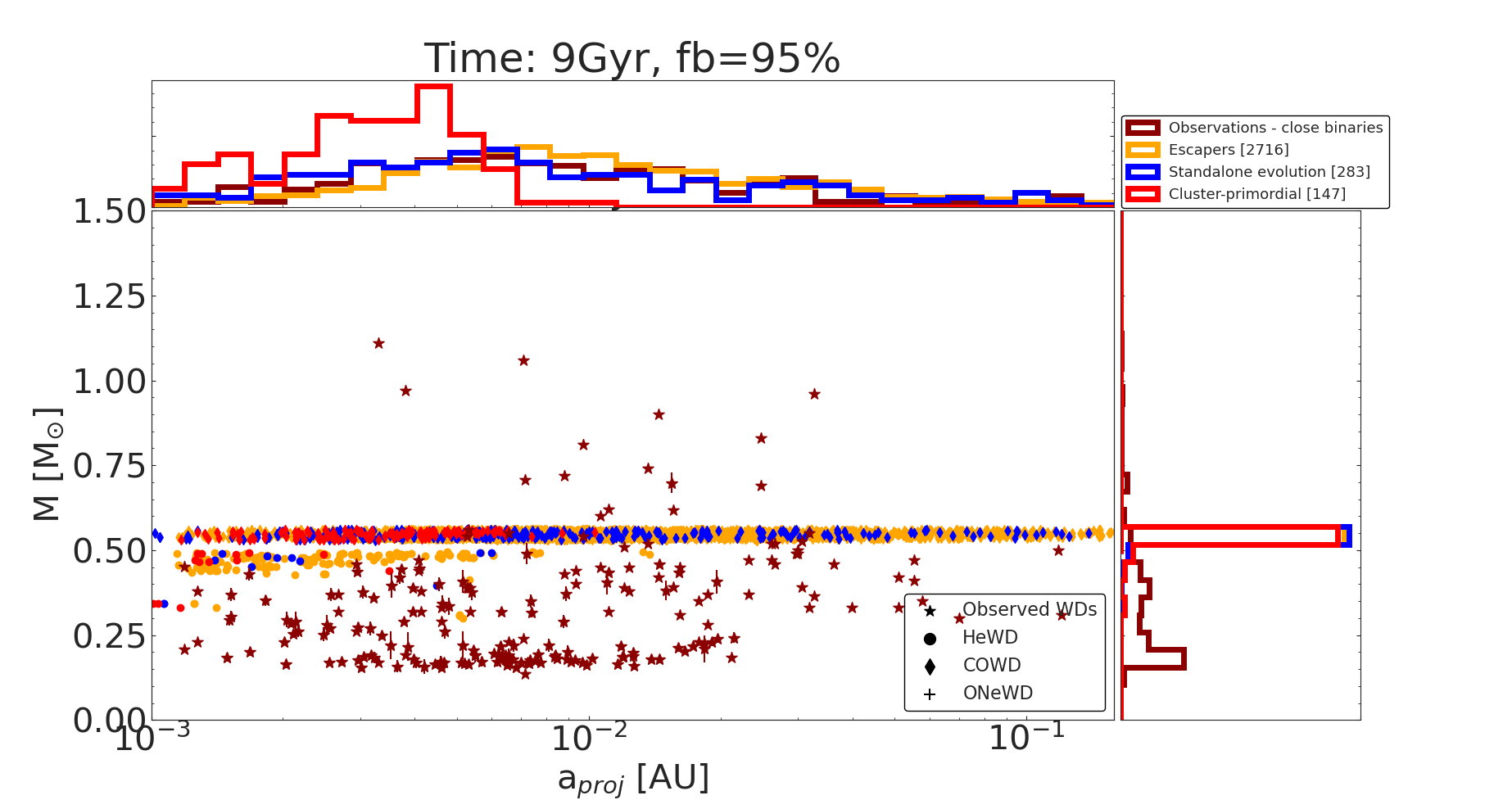}
    \end{minipage}
    \hspace{1em}  
    \begin{minipage}[b]{5cm}
        \caption{Projected separation plotted against mass of the lower mass WD in the binary for close binaries inside clusters with a 95\% initial binary fraction. See Fig.~\ref{fig:aMass_010} for an explanation of panels, colors, and markers}
        \label{fig:perMass_fb0.95}
    \end{minipage}
\end{figure*}

For a 95\% initial binary fraction (Fig.~\ref{fig:perMass_fb0.95}) we have a similar scenario, but there are small differences. At 2 Gyr we see a larger shift toward higher separations compared to a 10\% binary fraction. As expected, similarly to the  10\% binary fraction, we are not able to form the ELMs WDs that have been observed. At 5 and 9 Gyr, we have good agreement in separation between escapers, isolated binaries, and observations. In-cluster binaries are biased toward lower separations.

Looking at the number of close binaries we can see that the number of binaries is increasing with time for escapers and isolated binaries since over time more binaries   escape their host cluster or the components of the binary  evolve into WDs. However, for in-cluster binaries, the number is decreasing due to escaping binaries and dynamical interactions causing break-ups, mergers, or exchanges.

An interesting feature is a triangular  gap in Fig.~\ref{fig:perMass_fb0.95}, panel (a). This gap occurs between 0.02 and 0.07 AU in projected separation and 0.8 and 1.1 \msun in mass. This gap is present for all datasets and the reason for it is  currently not completely understood. However, it is most likely due to the pairing of objects in binaries when the initial cluster model is set up \citep{kiminki2012, sana2012, kobulnicky2014}.

\subsection{Wide binaries}
\label{sec:results-wide}
In this section we report our results involving wide binaries that were obtained by filtering our \mocca data according to \ref{sec:wideFiltering}. We split our results into clusters with initial binary fractions of 10\% and 95\%. As we discussed in Sect.~\ref{sec:results-close}, our limit in magnitude causes high mass WDs to become undetectable at later snapshots.

\subsubsection{Binary fraction of 10\%}
\begin{figure*}[ht]
    \centering
    \begin{minipage}[b]{12cm} 
        \centering
        \includegraphics[width=\textwidth]{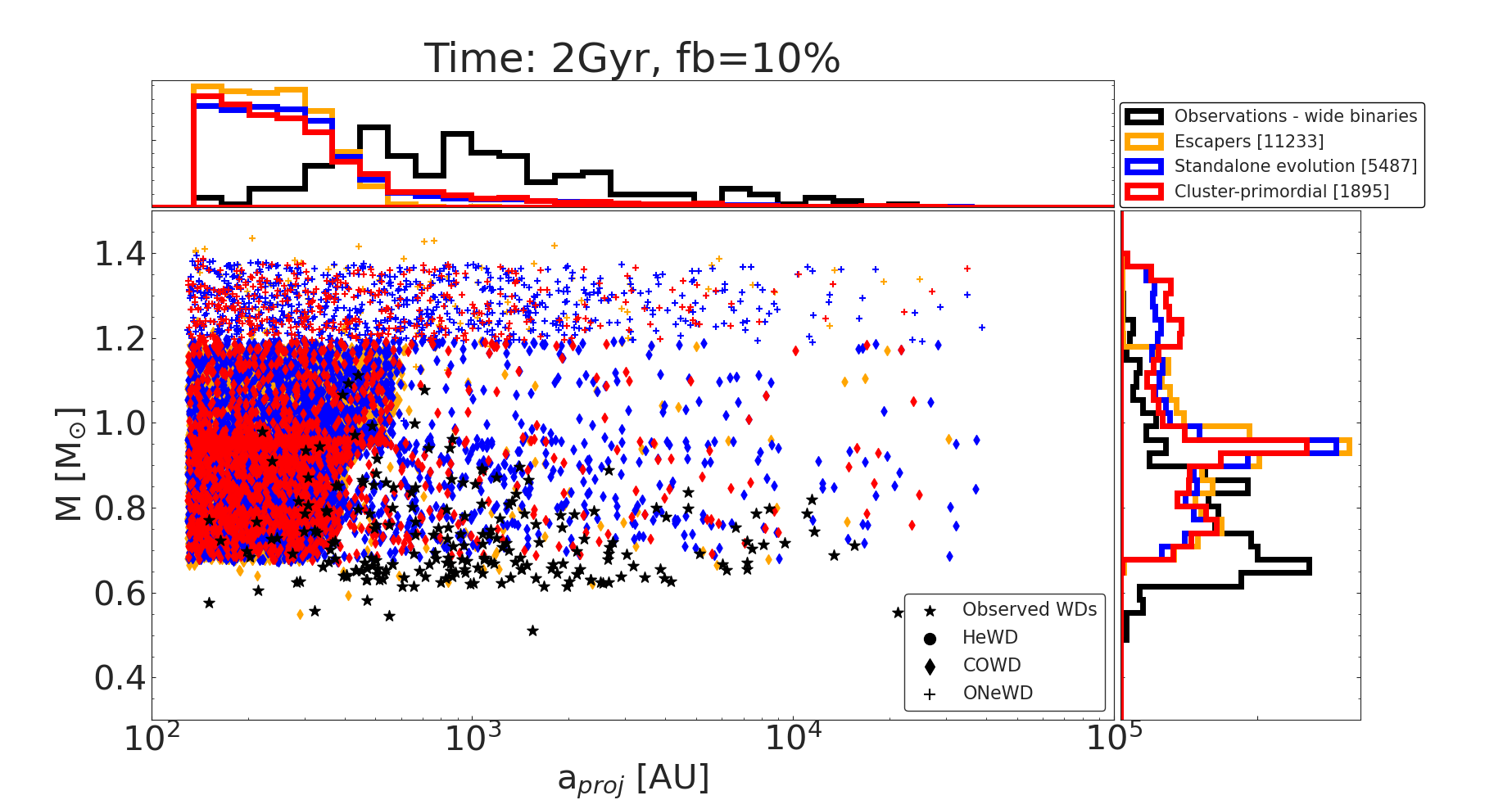}\\[2ex]
        \includegraphics[width=\textwidth]{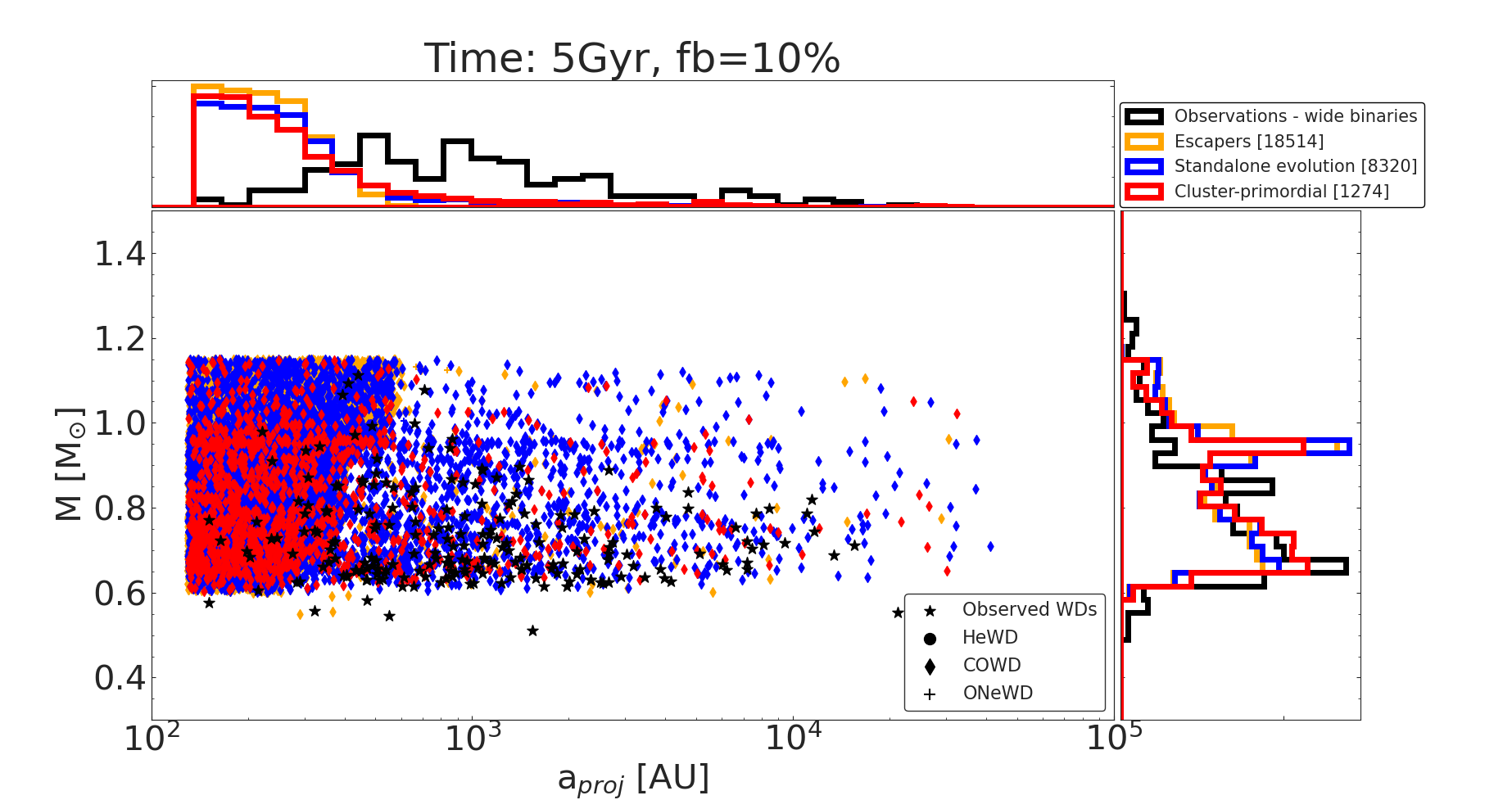}\\[2ex]
        \includegraphics[width=\textwidth]{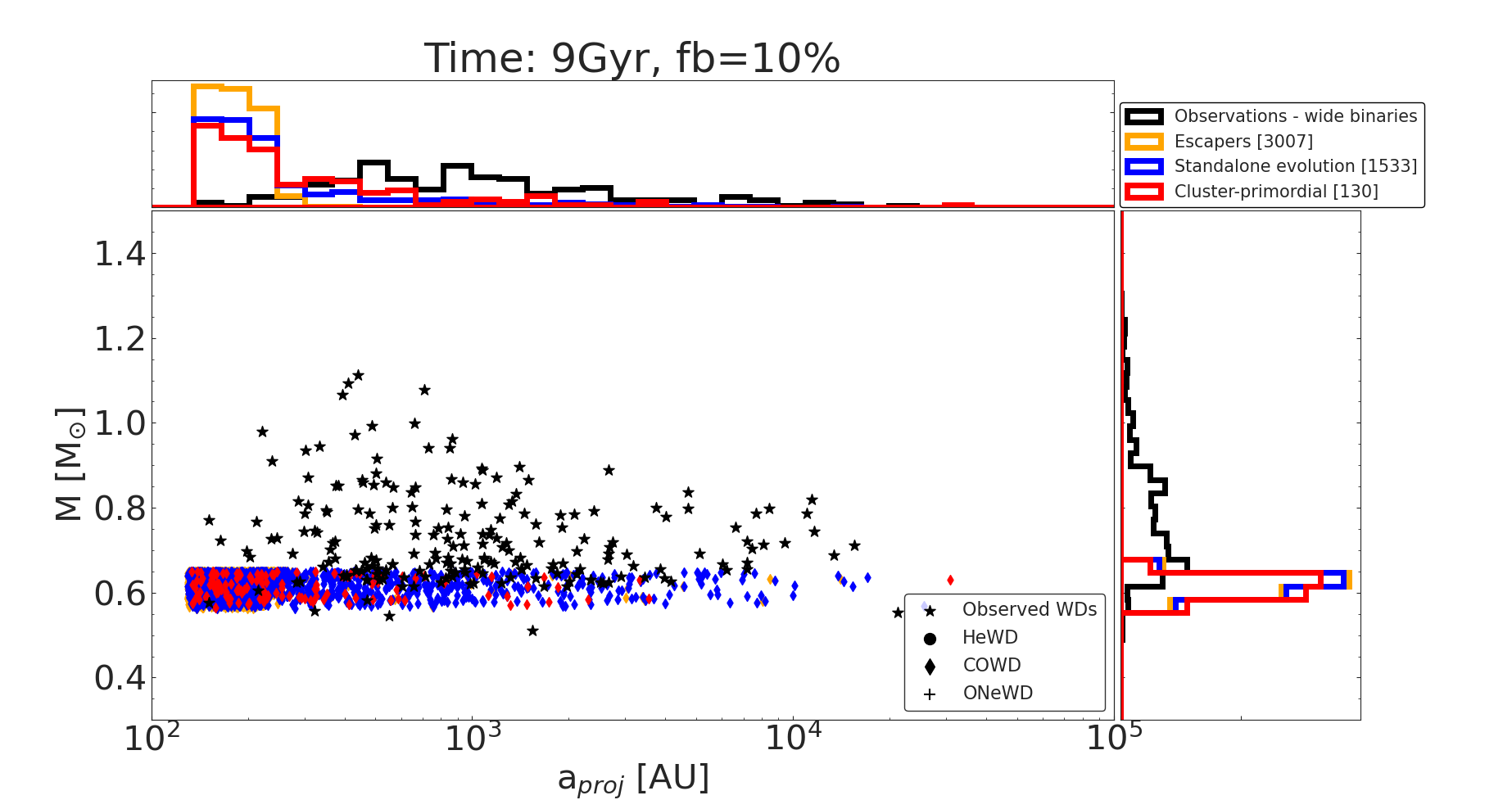}
    \end{minipage}
    \hspace{1em}
    \begin{minipage}[b]{5cm}
        \caption{Projected separation plotted against mass for wide binaries with a 10\% initial binary fraction. See Fig.~\ref{fig:aMass_010} for an explanation of panels, colors, and markers.}
        \label{fig:projSep_fb0.1}
    \end{minipage}
\end{figure*}

Figure \ref{fig:projSep_fb0.1} shows the distribution of projected separations plotted against mass of both binary components for a 10\% initial binary fraction at 2, 5, and 9 Gyr. For all snapshots, the binaries from \mocca are not as wide as the observed binaries. This occurs when we use a 10\% binary fraction; we have a cutoff at 100 au for initial semi-major axis. This is done under the assumption that the soft components of the binary populations are entirely gone and the remaining binaries are hard binaries. However, recent studies have shown that having nearly 100\% initial binaries with most being soft is a better assumption \citep[e.g.,][]{leigh2015,belloni2019}.

It is therefore unlikely that the binaries will evolve into larger separations. Since the absence of wide systems is seen in all three datasets, including the standalone evolution set, this cannot be due to dynamical interactions. Looking at the mass distribution we can see,
as expected, that the masses from our simulations depend on time. Over time, lower mass WDs binaries are   formed as expected, so for later times the distribution of masses becomes lower, in addition to this, the cooling process causes high mass WDs to be undetectable at later times. We find that at 5 Gyr our distribution aligns fairly well with the data from \cite{heintz2022}. At 2 Gyr the masses are biased toward higher masses, while at 9 Gyr they are biased toward lower masses due to the cooling of WDs.

\subsubsection{Binary fraction of 95\%}
\begin{figure*}[ht]
    \centering
    \begin{minipage}[b]{12cm}
        \centering
        \includegraphics[width=\textwidth]{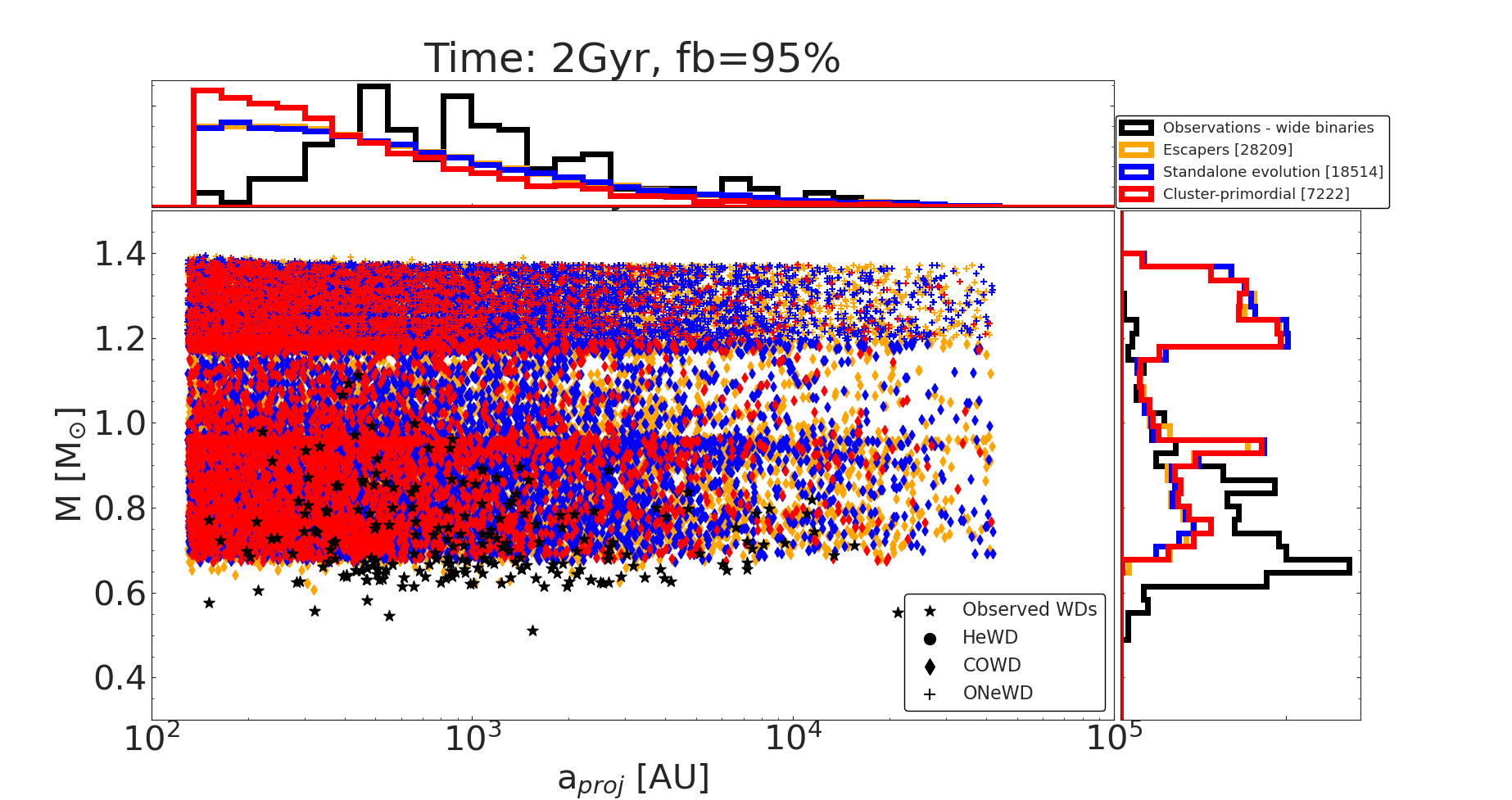}\\[2ex]
        \includegraphics[width=\textwidth]{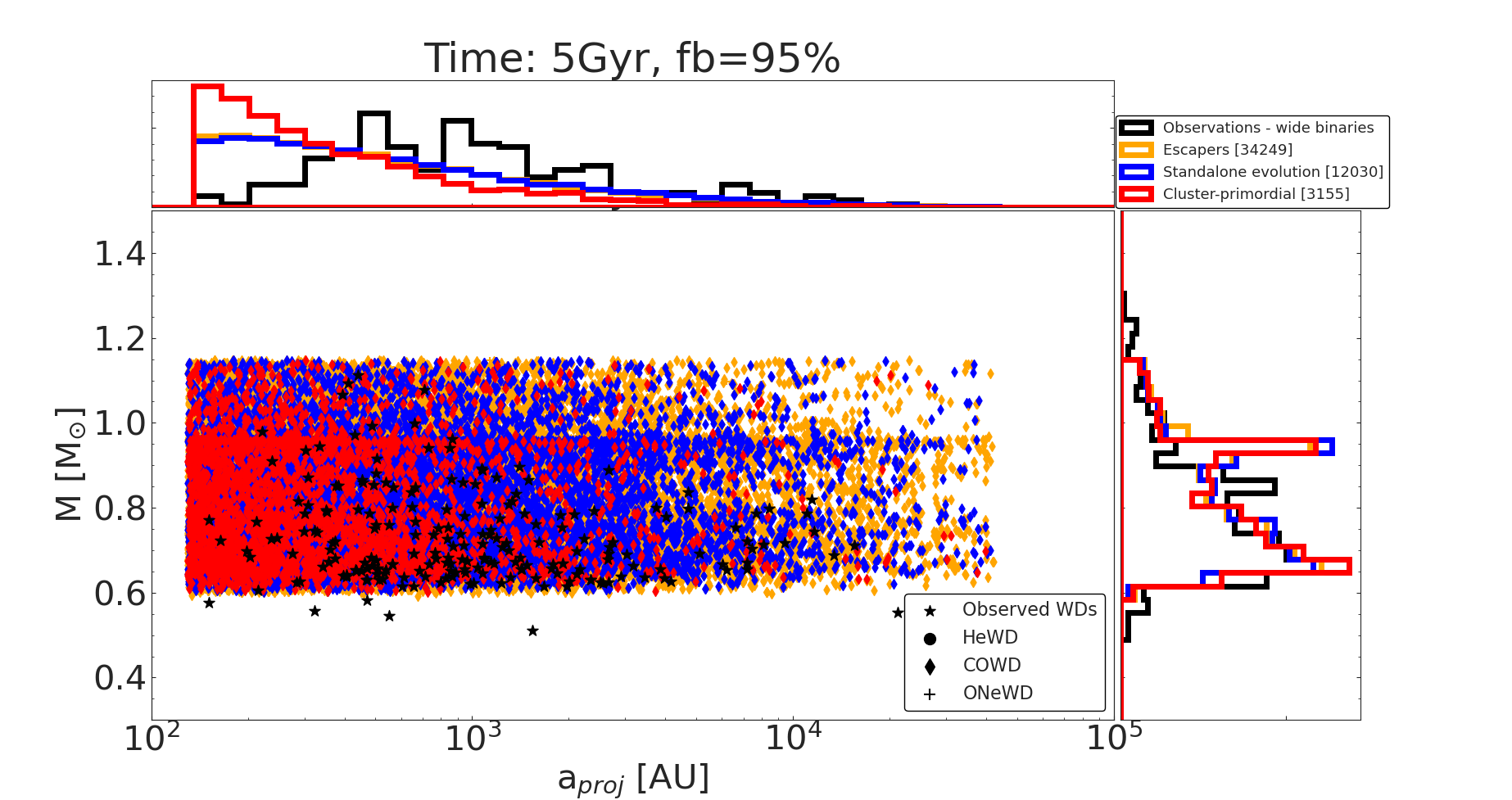}\\[2ex]
        \includegraphics[width=\textwidth]{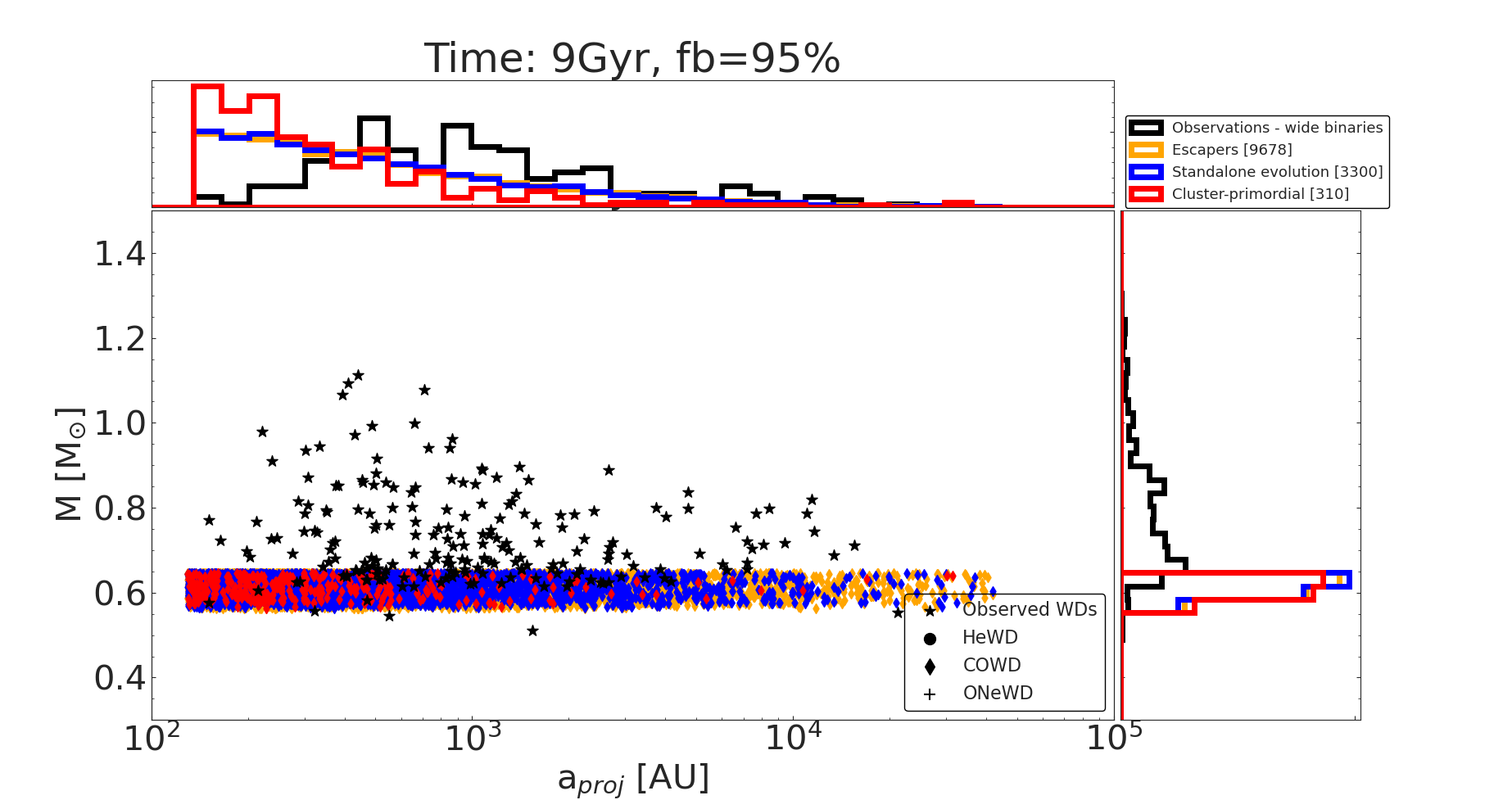}
    \end{minipage}
    \hspace{1em}
    \begin{minipage}[b]{5cm}
        \caption{Projected separation plotted against mass for wide binaries with a 95\% initial binary fraction. See Fig.~\ref{fig:aMass_010} for an explanation of panels, colors, and markers.}
        \label{fig:projSep_fb0.95}
    \end{minipage}
\end{figure*}

When using a  95\% binary fraction, we allow wider binaries to be formed initially. This causes our separation distribution to be more shifted toward wider binaries. This is shown in Fig.~\ref{fig:projSep_fb0.95}. All three datasets from \mocca are very similar for all snapshots, but we  point out the small differences. At 2 Gyr we have a tail extending toward wider binaries, but our distribution is still biased toward lower separations than the observations. The masses do not align very well at this snapshot: they are too high and the observed WDs have noticeably lower masses. Escapers and isolated binaries are very similar in their separation distribution, while in-cluster binaries have slightly smaller separations, which could be due to disruptions in dynamical interactions. 

At 5 Gyr the separation distribution is very similar to the 2 Gyr snapshot. However, the masses agree much better with observations. Due to more time passing, lower mass WDs can form. In addition, the cooling sequence removes high mass WDs from our sample. This causes a good agreement with observations in mass. At 9 Gyr, the separation distribution does not change in any significant way. However, the cooling process has now made the higher mass WDs undetectable and we are left with a small subset of WDs with mass around $0.6$ M$_{\odot}$. Similarly to the close binaries, we can see that over time the  number of binaries increases in the escapers and standalone evolution datasets, while the number decreases in the cluster dataset.

\subsubsection{Heintz age distribution}
\label{sec:results_WideHeintzAges}

\begin{figure*}
    \sidecaption
    \includegraphics[width=12cm]{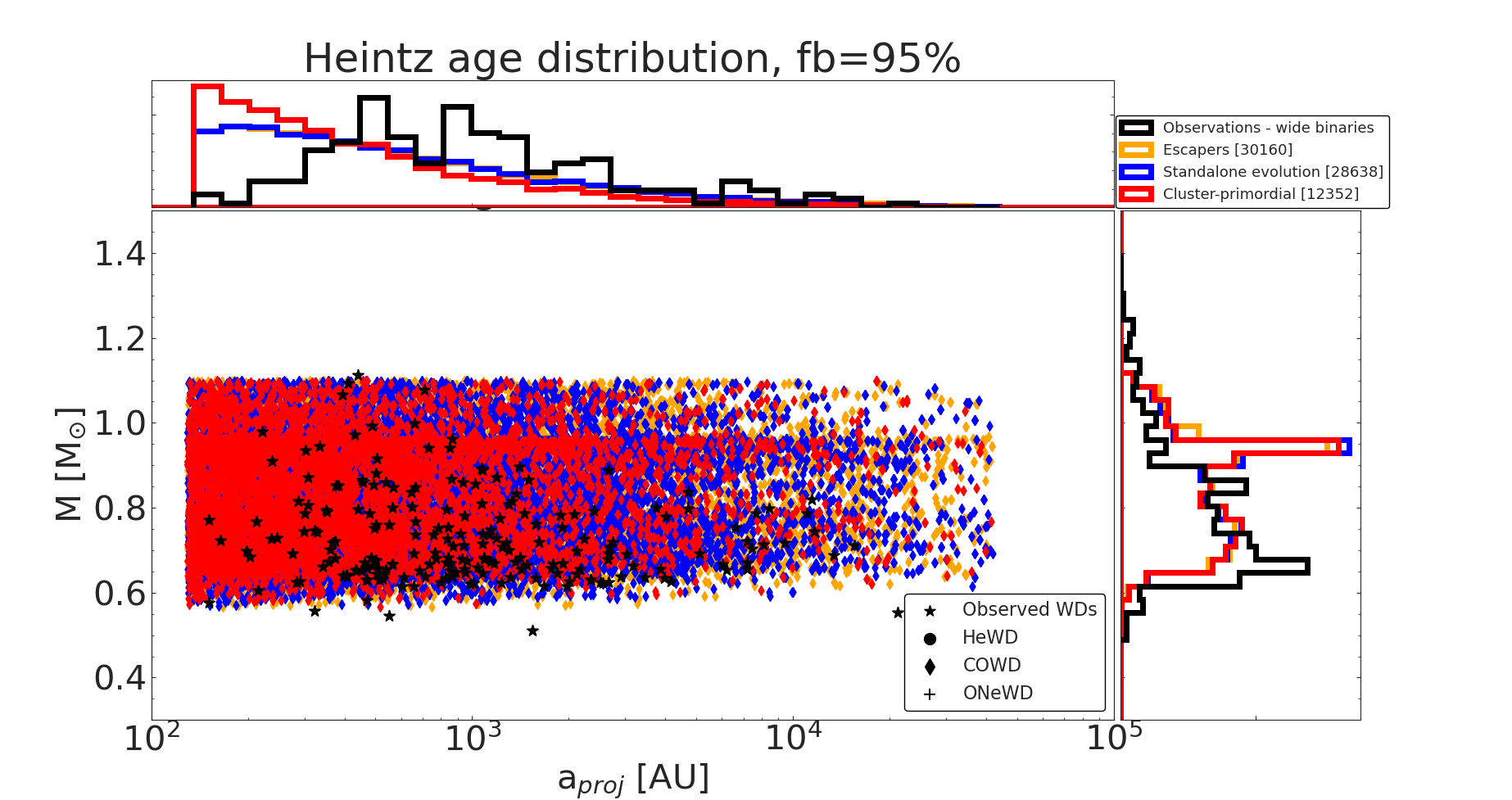}
    \caption{Projected semi-major axis plotted against mass of the WDs for our set with an age distribution similar to \cite{heintz2022}. See Fig.~\ref{fig:aMass_010} for an explanation of colors and markers.}
    \label{fig:projSep_heintz}
\end{figure*}

We can see in the previous plots that there is a time dependence in the mass distribution, which is the result of  lower mass WDs being formed and of higher mass WDs not being detectable due to the cooling process. Figure \ref{fig:projSep_heintz} shows a sample of our DWDs with the same age distribution as in \cite{heintz2022}. This is explained in Sect.~\ref{sec:wideFiltering}. We can see that by doing this we indeed get a fairly good mass distribution; our masses are only slightly lower, but there is a bias and a peak at around $0.9$ M$_{\odot}$. This peak is connected to \bse, and is in more detail than the envelope type. \bse handles both convective and radiative envelopes, but, as seen in this figure, the switch between them may be too abrupt to be realistic. Nevertheless, if we   disregard this peak, our mass distribution agrees very well with the observations.

\begin{figure*}
    \sidecaption
    \includegraphics[width=12cm]{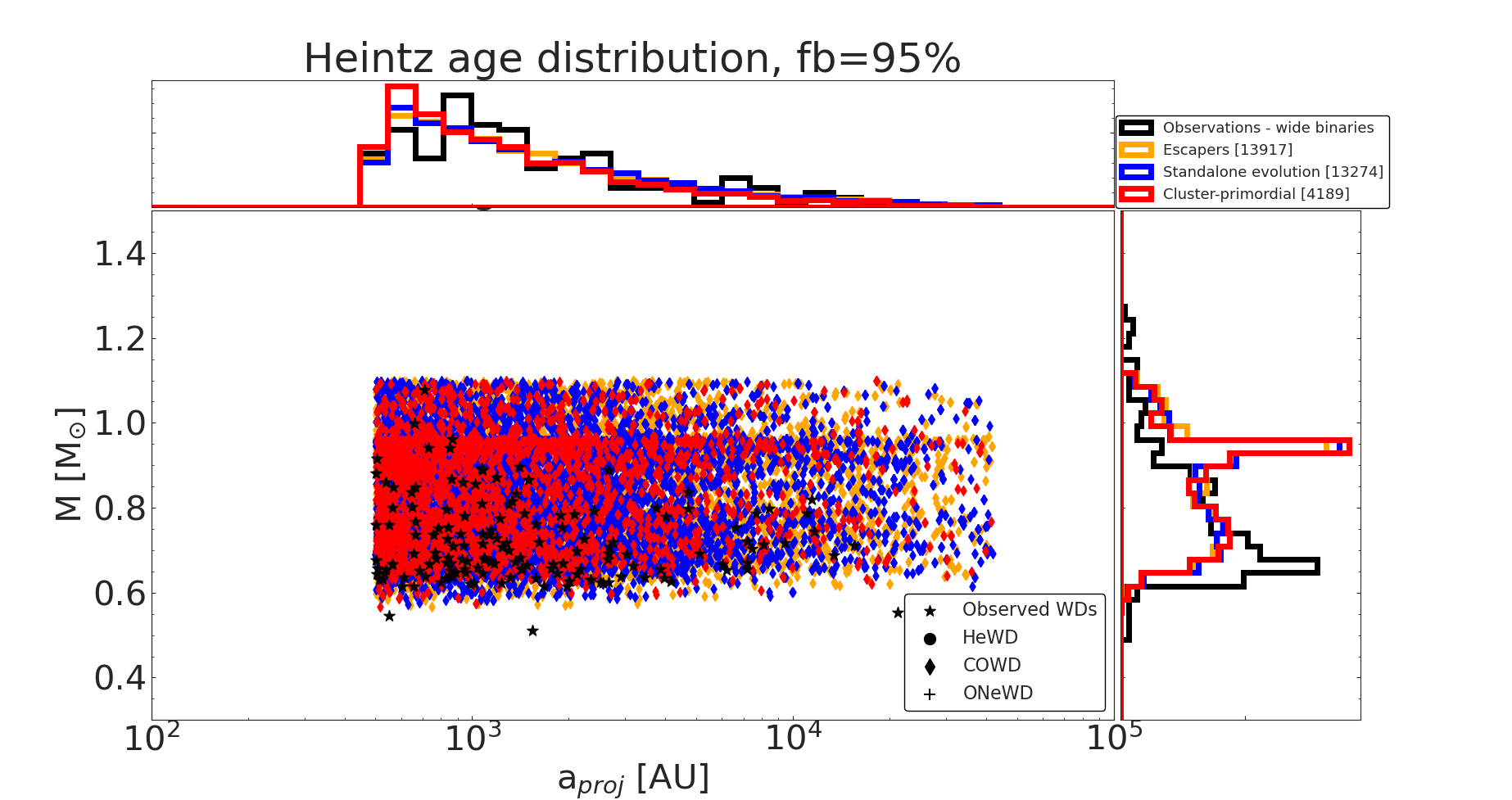}
    \caption{Projected separation plotted against mass of the WDs in the binary. In this figure we added an additional cutoff at 500 au to remove the tail of  smaller separation binaries found in the observational data. We see good agreement in both mass and separation. See Fig.~\ref{fig:aMass_010} for an explanation of colors and markers.}
    \label{fig:projSep_heintz_onlyWide}
\end{figure*}

However, this does not solve the problem of too small separations. This is most likely due to observational bias and WD binaries with smaller separations are not observed or included in this observational survey. While the observations have a slight bell curve shape with a peak around 800 au, the \mocca data extends to much smaller separations, and we are only looking at the far end tail. Due to this, we added an additional cut and only include binaries with projected separation above 500 au. This can be seen in Fig.~\ref{fig:projSep_heintz_onlyWide} where our agreement between separations is much better. With these filtering procedures we can reproduce the observations in a very good way, in both projected separation and mass for all three of our simulated datasets.

\section{Summary and conclusions}
\label{sec:sumAndConc}
In this paper we took a large number of DWDs from 197 \mocca simulations and compared them to observed DWDs. This proved to be difficult due to strong observational biases. However, with some filtering and cuts on the \mocca data, we managed to get good agreement in separation for close binaries and for both separation and masses for wide binaries. One important point about our data is that we   incorporated multiple stellar populations in our initial model. In these models we have a first population that is tidally filling and not too concentrated and a second population that is significantly more concentrated than the first population. Thus, our initial setup differs greatly from the traditional approach of having one dense initial population. In these new setups we find a much larger number of escapers compared to simulations using the traditional one-population setup. We saw that many clusters lose 30-40\% of their mass in just the first few Myr due to objects escaping the cluster induced by stellar evolution mass loss. Due to the early escape of these objects, they are also mostly unaffected (or in many cases, completely unaffected)  by dynamics before their escape. We could see this by the similarities in results between the standalone evolution dataset and escapers dataset. This means that DWDs born in the field and those born inside GCs and later ejected are, on average, similar.

Our GC initial conditions gave us both tidally filling and underfilling clusters. Since tidally filling clusters   lose more mass in the early stages of   evolution, the number of escaping DWDs was expected to be different for these two different types as well. We find that from tidally filling clusters, there are approximately two times more DWD escapers at 2 Gyr and approximately 1.6 times more DWD escapers at 9 Gyr compared to tidally underfilling clusters. However, we do not find any significant differences in the statistical distribution of separations of DWDs that have escaped from tidally filling and underfilling clusters.

\begin{figure}
  \resizebox{\hsize}{!}{\includegraphics{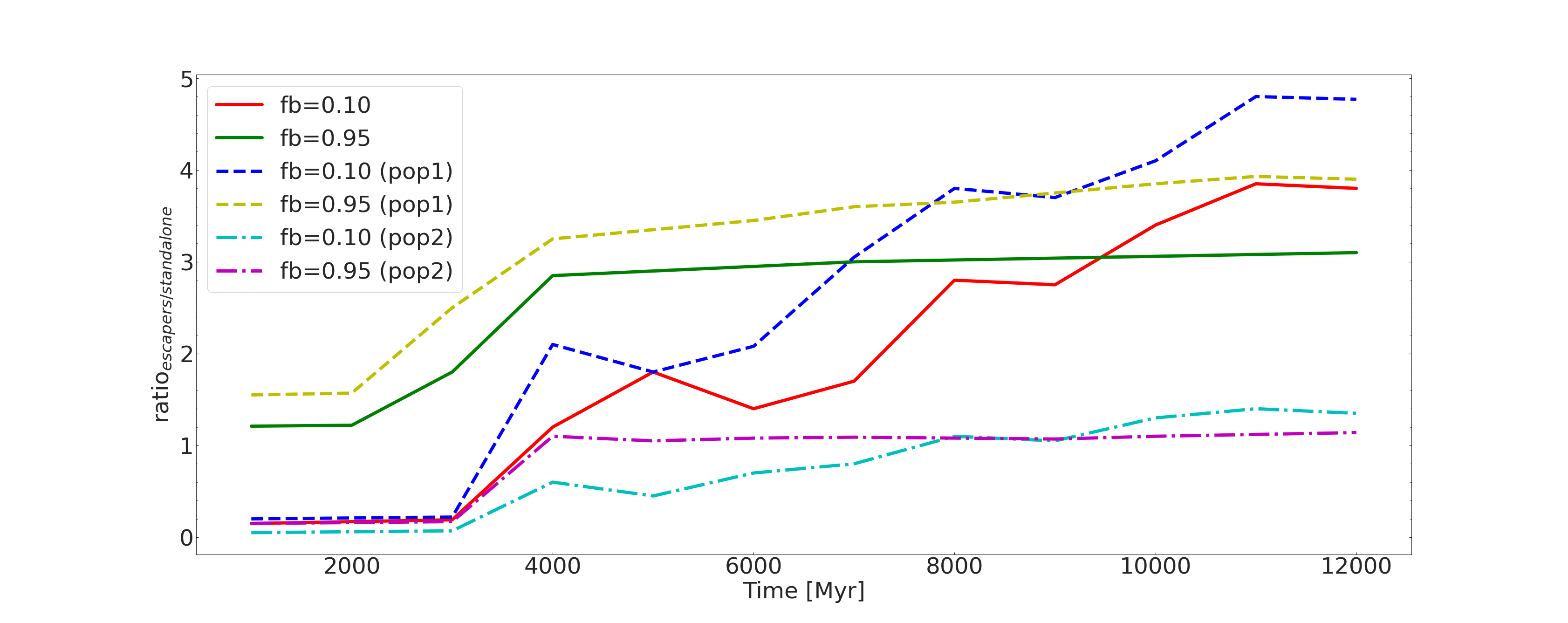}}
    \caption{Evolution of ratio of    escaping DWDs to standalone DWDs for every Gyr between 1 and 12 Gyr. We have six groups in this figure. The solid lines show both populations together; the dashed lines show population 1 only; and the dotted lines show population 2 only.}
    \label{fig:rat_escStand}
\end{figure}

A significant finding is related to the number of DWDs in each set. We found that for most clusters and snapshots, the number of DWDs in the escaping dataset is far larger than the other two sets (in-cluster and standalone binaries). This is clearly shown in Fig. \ref{fig:rat_escStand} where we plot the ratio of the number of escaping DWDs to standalone DWDs. For a  95\% binary fraction, we have slightly more escaping DWDs at 1 Gyr than standalone DWDs; however, at later times this ratio grows and we see a faster increase in the number of escaping DWDs. At 12 Gyr, there are approximately three times more escaping DWDs than standalone ones. For the  10\% binary fraction, we start with significantly fewer escaping DWDs. At 3 Gyr we have approximately the same number, and at later times we have more escaping than standalone. At 12 Gyr, we have almost four times more escaping DWDs than standalone. This indicates that even though many binaries escape early, they are still influenced by dynamics in a way that causes them to be more likely to form WDs within a Hubble time. This is true even for less dense clusters and clusters with a low binary fraction. For populations 1 and 2 separately, we can see that the big difference in this ratio comes from population 1, especially for the 95\% binary fraction at early times. 

\begin{figure}
    \subfigure[]{\includegraphics[width=0.5\textwidth]{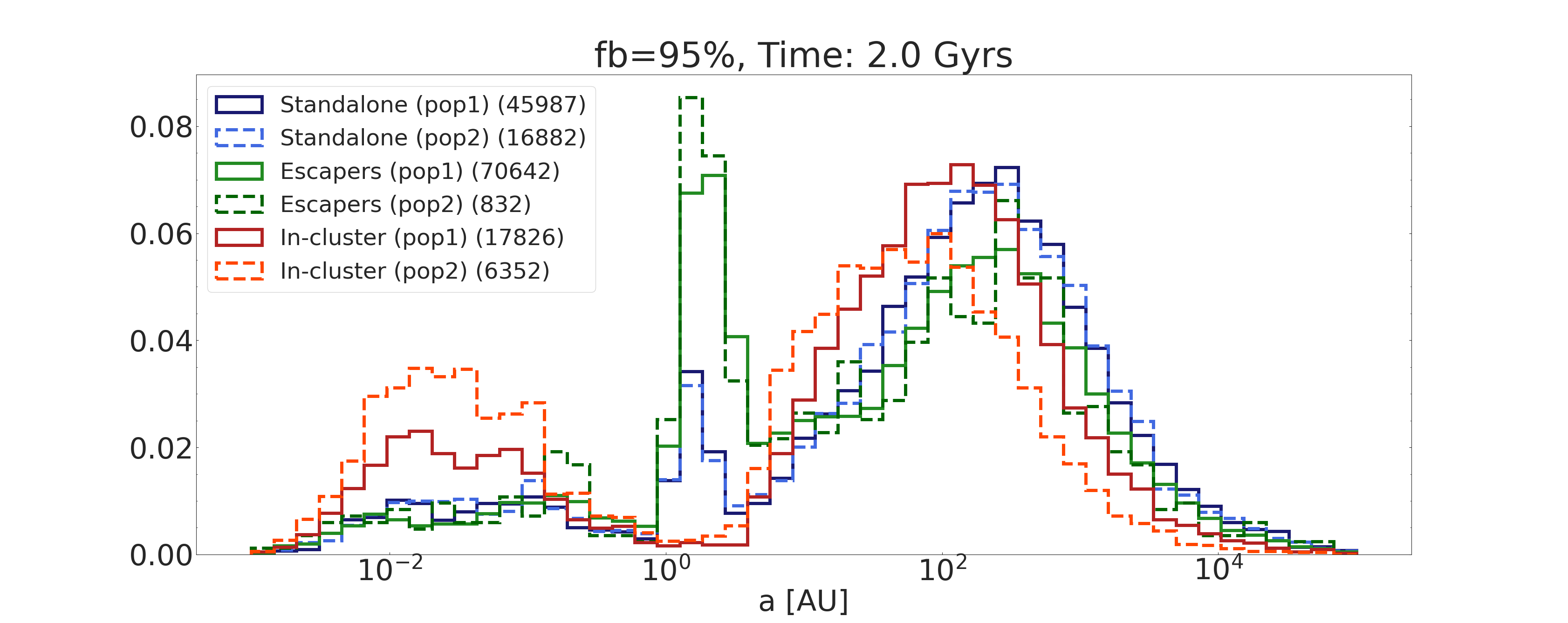}} 
    \subfigure[]{\includegraphics[width=0.5\textwidth]{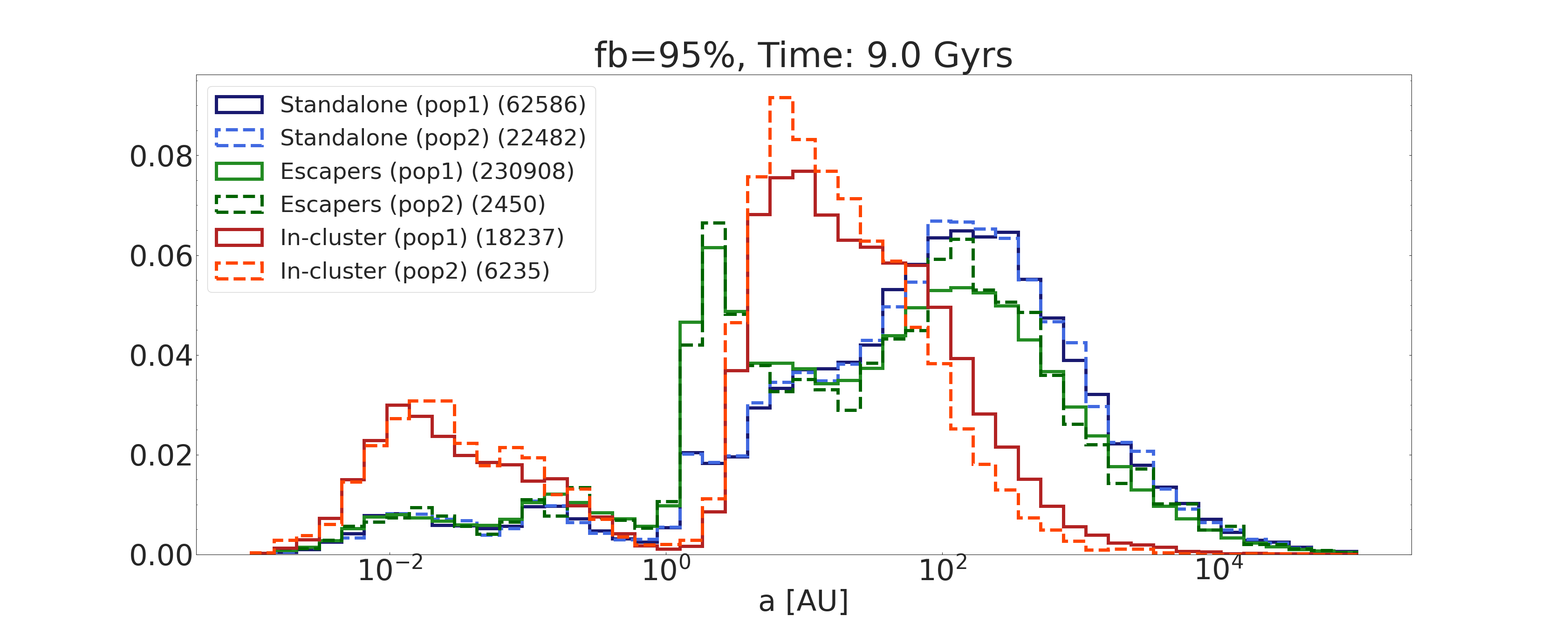}}
    \caption{Histogram of the semi-major axis distribution for the three \mocca datasets. Populations 1 and 2 are split into different groups from a snapshot at 9 Gyr for the 95\% binary fraction.}
    \label{fig:gap_hist}
\end{figure}

When looking at the whole range of semi-major axes (see Fig.~\ref{fig:aMass_010}), we can see a gap in the in-cluster population between $\sim$0.5 and $\sim$25 au. For the escape and standalone populations, we do not see this clear gap. This can be seen more clearly in Fig.~\ref{fig:gap_hist} where we plot a histogram of the distribution of semi-major axes for the three datasets from \mocca at 2 Gyr (panel a) and 9 Gyr (panel b) for the 95\% binary fraction. We also split populations 1 and 2 into separate groups in order to see the effect of the two populations. In the plot we clearly see the gap in the in-cluster set for both population 1 (dark red) and population 2 (orange) at both times. Comparing the two populations at 2 Gyr, for small separations we have an approximately 55\% higher fraction of population 2 binaries than population 1 binaries. However, at larger separations, we have an approximately 25\% higher fraction of population 1 binaries than population 2 binaries. At 9 Gyr the two populations have evened out and the differences are much smaller.

This gap is nonexistent for the escapers and standalone datasets. Instead, there is a sharp increase of $\sim$1 au. This big peak is mostly visible for escapers at 2 Gyr, but is there for both datasets at both snapshots. This very big population of escaping binaries at $\sim$1 AU seems to indicate that this gap is partly due to escaping binaries; however, the reason why these particular binaries   escape is still uncertain. The fact that we   see a similar peak, although not as large, for the standalone runs seems to indicate that we have some bias in our initial datasets to create WD binaries with these separations. Combining this with a large mass loss in the first few Myr can lead to features like this. However, this our aim was to compare our results to observations, and finding a more conclusive answer to this gap is beyond the scope of this paper. We can also note that, from these plots, there are no significant differences in populations 1 and 2 for the escapers and standalone sets. Even though there is no statistical difference in the distribution of separation between populations 1 and 2, there is a clear number difference. We have almost 100 times more escapers from population 1 than 2, at both 2 and 9 Gyr. This causes population 1 to provide the large bulk of escaping binaries when analyzing the data as a whole.

Our main conclusions can be summarized as follows:
\begin{itemize}
    \item Observations of DWD systems are biased and limited, which make direct comparisons impossible without filtering \mocca data.
    \item When using a 10\% binary fraction, we are not able to reproduce the separation distribution of wide binaries. This is due to how the initial distribution of semi-major axis is set up for a 10\% initial binary fraction cluster (see Fig.~\ref{fig:projSep_fb0.1}).
    \item Using a 95\% binary fraction with the initial semi-major axis distribution given by \cite{belloni2017} allows us to form wider binaries that agree better with observations (see Fig.~\ref{fig:projSep_fb0.95}).
    \item Using an   age distribution similar to that used in \cite{heintz2022} and adding a lower limit to the separation causes our 95\% binary fraction dataset to agree very well with observations in both separation and mass (see Fig.~\ref{fig:projSep_heintz_onlyWide}).
    \item The masses of WDs in wide binaries agree well with observations at 5 Gyr, but are too high at 2 Gyr and too low at 9 Gyr. This is true for both the 10\% and 95\% binary fractions.
    \item For close binaries, our data agrees well with observed binaries in separation, in particular for escapers and isolated binaries;   the in-cluster binaries also agree fairly well (see Figs.~\ref{fig:perMass_fb0.1} and \ref{fig:perMass_fb0.95}).
    \item The very low masses of ELM WDs are not able to be reproduced with \bse. Uncertainties in binary evolution related to magnetic braking, mass stability criteria, and mass transfer modeling make it impossible to form these extremely low mass WDs.
    
    \item Due to the addition of multi-stellar populations and a less dense first population, a large amount of objects are removed from the clusters in the first few Myr. This leads to a much larger number of escapers than in-cluster binaries and standalone evolution binaries. The main difference between the escapers and standalone binaries is that the escapers have the possibility to be hardened in dynamical interactions before their escape. This can cause wide binaries to form DWDs on a shorter timescale, thus increasing the number of escaping DWDs. 
    
    \item The cooling sequence of WDs and the magnitude limits of observations indicate the mostly young WDs are being observed. This is confirmed by the age estimates from \cite{heintz2022}, where a majority of binaries have an age younger than 3 Gyr. We are hoping that this conclusion   holds for open younger clusters, which are also losing members that are slightly affected by dynamics.

    \item There is room for future analysis on these subjects, in particular the gap that we found in the separations of DWDs. The focus on this paper was on comparisons to observations, and we were not able to find a concrete explanation to this gap. In addition, with future observational missions there will be more data available. This will lead to better comparisons with simulated data, especially since most observational data is constrained to either close or wide binaries.
\end{itemize}


\begin{acknowledgements}
We would like to thank the reviewer for providing comments and suggestions that helped to improve the quality of the manuscript.
LH, MG, AH, AA, GW were supported by the Polish National Science Center (NCN) through the grant UMO-2021/41/B/ST9/01191.
DB acknowledges financial support from {FONDECYT} grant number {3220167}.
AA acknowledges support for this paper from project No. 2021/43/P/ST9/03167 co-funded by the
Polish National Science Center (NCN) and the European Union Framework Programme for Research
and Innovation Horizon 2020 under the Marie Skłodowska-Curie grant agreement No.
945339.
For the purpose of Open Access, the authors have applied for a CC-BY public copyright license to any Author Accepted Manuscript (AAM) version arising from this submission.
\end{acknowledgements}


\section*{Data availability}
Input and output data for the globular cluster simulations carried out in this paper is available at \url{https://zenodo.org/records/10865904}

\bibliographystyle{aa}
\bibliography{ref.bib}

\end{document}